\documentclass[12pt,a4paper]{article}

\usepackage[utf8]{inputenc}
\usepackage{amsmath}
\usepackage{amsthm}
\usepackage{amsfonts}
\usepackage{amssymb}
\usepackage{amsthm}
\usepackage{tikz}
\usepackage[toc,page]{appendix}
\usepackage{multirow}
\usepackage{tikz}
\usepackage{lscape}
\usepackage[a4paper]{geometry}
\usepackage{longtable}
\usepackage{pstricks}
\usepackage{algorithm}
\usepackage{algorithmic}
\usepackage[square,sort,comma,numbers]{natbib}
\renewcommand{\mathbf}[1]{\textbf{\mathversion{bold}$#1$}}

\title{A General Hidden State Random Walk Model for Animal Movement \footnote{Accepted manuscript: \href{http://dx.doi.org/10.1016/j.csda.2016.07.009}{doi: 10.1016/j.csda.2016.07.009}}}

\newcommand\encircle[1]{%
  \tikz[baseline=(X.base)]
    \node (X) [draw, shape=circle, inner sep=0] {\strut #1};}

\date{}

\usepackage{authblk}

\usepackage{hyperref} 
\usepackage{helvet}

\author{A. Nicosia$^{1}$\thanks{aurelien.nicosia.1@ulaval.ca}}
\author{T. Duchesne$^{1}$\thanks{thierry.duchesne@mat.ulaval.ca }}
\author{L.-P. Rivest$^{1}$\thanks{louis-paul.rivest@mat.ulaval.ca}}
\author{D. Fortin$^{2}$\thanks{daniel.fortin@bio.ulaval.ca}}
\affil{$^{1}$D\'{e}partement de math\'{e}matiques et de statistique, Universit\'{e} Laval, \\
pavillon Alexandre-Vachon, 1045, av. de la Médecine, bureau 1056, Québec (Qc), G1V 0A6, Canada}
\affil{$^{2}$D\'{e}partement de biologie, Universit\'{e} Laval, 
\\
pavillon Alexandre-Vachon,
1045, av. de la Médecine, bureau 3058 Québec (Qc), G1V 0A6, Canada}

\begin{document}
\maketitle
\copyright  2017. This manuscript version is made available under the CC-BY-NC-ND 4.0 licence \href{https://creativecommons.org/licenses/by-nc-nd/4.0/}{https://creativecommons.org/licenses/by-nc-nd/4.0/}








\begin{abstract}
A general hidden state random walk model is proposed to describe the movement of an animal that takes into account movement taxis with respect to features of the environment. A circular-linear process models the direction and distance between two consecutive localizations of the animal. A hidden process structure accounts for the animal's change in movement behavior. The originality of the proposed approach is that several environmental targets can be included in the directional model. An EM algorithm that enables prediction of the hidden states of the process is devised to fit this model. An application to the analysis of the movement of caribou in Canada's boreal forest is presented.\end{abstract}



%



%

\newpage
   \section*{List of symbols}
\textbf{Data}
\begin{longtable}[l]{p{70pt}  p{350pt}}
$T$ & number of observed animal's locations.\\
  $y_t$		& direction between the animal's locations at time steps $t$ and $t+1$\\
  $y_{0:T}$ & set of all observed directions $y_0,\ldots,y_T$\\
  $d_t$		& distance between the animal's locations at time steps $t$ and $t+1$\\
    $d_{0:T}$ & set of all observed directions $d_0,\ldots,d_T$\\

  ${x}_{it}$		& value of $i$th explanatory angle variable\\
    ${z}_{it}$		& value of $i$th explanatory real variable\\
$\mathcal{F}_t^o$ & observed information: directions, distances and explanatory variables gathered form time $0$ up to time $t$ \\
   \end{longtable}
   \textbf{Hidden process}
\begin{longtable}[l]{p{70pt}  p{350pt}}
  $S_t$		&  state (behavior) in which the animal is at time step $t$\\
        $S_{kt}$		&  indicator function equal to $1$ if $S_t=k$ and $0$ otherwise \\
    $S_{0:T}$ & set of all states $S_0,\ldots,S_T$\\
$\mathcal{F}_t^c$ & complete information: information in $\mathcal{F}_t^o$ and the hidden states $S_0,\ldots,S_T$\\
$p(S_t|\mathcal{F}_{t-1}^c)$ & conditional probability mass of the hidden state $S_t$ \\
$\pi_{hk}$ & state transition probability $\mathbb{P}(S_t=k|S_{t-1}=h)$\\
   \end{longtable}

   \textbf{Observed trajectory}
\begin{longtable}[l]{p{70pt}  p{350pt}}
  $h(y_t,d_t|S_t,\mathcal{F}_{t-1}^c)$		& conditional joint density of the observed data $(y_t,d_t)$ \\
   $f(y_t|S_t,\mathcal{F}_{t-1}^c)$		& conditional density of the direction $y_t$\\
   $f_k$ &density of the direction $y_t$ given that the hidden state is $k$ at time $t$  \\
         $\kappa^{(k)}$ 	& vector of parameter of $f_k$ \\
                  $\mu_t^{(k)}$ 	& mean direction of the von Mises density $f_k$ \\
         $\ell_t^{(k)}$ 	& concentration parameters of the von Mises density $f_k$ \\

   $g(d_t|S_t,\mathcal{F}_{t-1}^c)$ 	& conditional density of the distance $d_t$\\
      $g_k$ 	& density of the distance $d_t$ given that the hidden state is $k$ at time $t$ \\
         $\lambda_1^{(k)},\lambda_2^{(k)}$ 	& shape and scale parameters of $g_k$\\  

   \end{longtable}

\newpage
\section{Introduction}
\label{s:intro}
In animal ecology, being able to understand and model the movement of animals is fundamental (\cite{Nathan2008}).
For example, animal behaviorists want to see to what extent animals have preferred movement directions or are attracted towards several environmental targets,
such as food-rich patches and previously visited locations (spatial memory effect) (\cite{Latombe2014}).
The development of Global Positioning System (GPS) technology permits the collection of a large amount of data on animal movement. This can be combined to data available from geographic information systems (GIS) to investigate how the environment influences animal displacement. To achieve this goal, robust statistical techniques and flexible animal movement models are required.

Discrete time models for animal movement are actively being developed and investigated (\cite{Holyoak2008}). Because displacement in discrete time
can be characterized by the distance and the direction between two consecutive localizations, circular-linear processes can be used to model movement in 2D.
A basic model is the biased correlated random walk (BCRW) of \cite{Turchin}; it predicts the next motion angle as a comprise between the current one (often called directional persistence if the bias is towards zero) and the direction towards a specific target (also called directional bias). 
Several authors have built models to adapt or generalize the BCRW so that it can be applied in different contexts. \cite{Jonsen2005} directly model the $(x,y)$-coordinates at a given time-step as a function of coordinates at the previous time-step with bivariate normal distributions to deal with data acquired at irregular time intervals. An alternative formulation of this model is proposed by \cite{Shimatani2012}, but this formulation of the BCRW does not allow to include multiple directional biases.
 Despite the rapid increase in the development of movement analysis, most quantitative techniques still consider only two directional targets when estimating the mean direction of BCRWs. However, habitat selection studies demonstrate that animal movement can be influenced simultaneously by more than one or two environmental features (\cite{Moreau2012}).
Our first generalization of the BCRW is therefore to make the mean direction of the process depend upon several directional targets. To do so, we embed the directional model recently proposed by 
\cite{Rivest2015} within the BCRW and show how it is easily interpretable. We also show that its estimation is numerically more stable than other BCRW models.

Often, the movement trajectory of animals involves multiple movement states or behaviors (\cite{Fryxell2008}). For instance, in their analysis of bison movement, \cite{Langrock2012} identified two states, ``exploratory" and ``encamped". The former has long traveled distances and turning angles between two consecutive locations that tend to be concentrated around zero, while the latter one is characterized by short distances and almost uniformly distributed turning angles. Multiple movement behaviors can be accounted for by introducing hidden states in the models.
\cite{baum66} give a general presentation of these models and \cite{Morales2004}, \cite{Jonsen2005}, \cite{Holzmann2006} and \cite{Langrock2012} are examples of the use of hidden state models to analyze angular-distance data in ecological applications. The second main contribution of our work is to introduce more flexible hidden state models that can accommodate directional persistence as well as the simultaneous influence of several environmental targets that can vary from state to state.
Further, by using the EM algorithm to fit the model, we are able to compute the posterior probabilities of the hidden state for each step of the animal's trajectory. Because these probabilities take into account the targets that are important in each hidden state, they can be used to understand the relative roles of these individual targets on the overall movement and space-use patterns of individuals. They can also serve as input values in movement simulations, such as individual-based movement models (e.g., \cite{Latombe2014}). Finally, these probabilities can highlight some regions in the landscape to be identified as patches of interest.

 The proposed model for animal motion data, a multi-state  circular-linear process, is introduced in Section 2. Each state has its own angular regression model featuring several environmental targets and directional persistence as introduced in \cite{Rivest2015}. The new model is not a Hidden Markov Model (HMM); it belongs to a wider class called switching Markov models investigated in \cite{Rydn2004} and \cite{Sylvia13}.  Its parameters are shown to be identifiable. We use the EM algorithm to maximize the likelihood, using a filtering smoothing algorithm (see  \cite{Sylvia13}), to carry out the expectation step; details are given in Section
 3. Section 4 investigates the finite sample properties of the estimators by simulation. Section 5 shows an application of the method to the analysis of the movement of caribou in the Cote-Nord
 region of Quebec, Canada. Section 6 concludes the paper with a discussion.
\section{A General Multi-State Random Walk Model}
\label{s:model}
Let us suppose that we follow an animal equipped with a GPS collar which provides the animal location at regular time intervals, for example every 4 hours.
Additional geographic information about multiple habitat features is available.
The data set consists of the time series
\begin{equation}\label{obsdata}
\left\lbrace \left(y_t,d_t,\mathbf{x}_t,\mathbf{z}_t\right),t=0,\dots,T \right\rbrace,
\end{equation}
where $y_t \in [0,2 \pi)$ and $d_t \geq 0$ represent the direction (bearing) and the distance, respectively, between the animal's location at time step $t$ and time
 step $t+1$ and $\mathbf{x}_t=(x_{1t},\dots,x_{pt})$ (resp. $\mathbf{z}_t=(z_{1t},\dots,z_{pt})$ ) are the values of $p$ explanatory angular (resp. real) variables measured
 that are potentially useful to predict $y_t$ or $d_t$.
 The explanatory variables $x_{it}, z_{it},i=1,\dots,p $ are associated to the directions to and the distances from these targets with respect to the position of the animal at time $t-1$. Explanatory variable $z_{it}$ can also be an indicator variable, see the \ref{a:D}.
For instance in the application of Section 5, $\mathbf{x}_t=(x_{1t},x_{2t})$ are the angles of the directions to the closest regenerating wood cut and the direction to the closest ``patch'' visited by the animal in the past, respectively. We also denote the set of all observed directions and distances by $(y_{0:T},d_{0:T})=\left\lbrace (y_t,d_t)  ,t=0,\dots,T \right\rbrace$.
We often need to condition on all the information (directions, distances, explanatory variables) gathered from time 0 up to time $t$; we denote this information by the filtration
$\mathcal{F}_t^o$. Our goal is to develop a suitable model for this type of data, along with the associated inference procedures.
\subsection{A General Hidden State Model}
Animals tend to adopt different movement behaviors at different times (\cite{Fryxell2008}). Clearly, such a change in behavior implies a change in the distribution of the values of the observed directions and distances. Because the animal's behavioral state over time is unobserved, we consider here a hidden process ${S}_t$, with $t=0,\dots,T$, that represents the state (behavior) in which the animal is at time step $t$. We denote by $\lbrace 1,\dots,K \rbrace$ the set of possible states of $S_t$, we put $S_{0:T}=\{S_0,\ldots,S_T\}$.
Conceptually, it is useful to define quantities that depend on both the observed and unobserved data. To this end, we define
the complete data filtration $\mathcal{F}_t^c$ as the filtration generated by the observed data filtration $\mathcal{F}_t^o$ and the hidden information up to time $t$.

The joint density of the complete data is

\begin{equation}
\label{Modgene}
f(y_{0:T},d_{0:T},S_{0:T})= \prod_{t=1}^T p \left( S_t | \mathcal{F}_{t-1}^c\right) h \left(y_t,d_t|S_t , \mathcal{F}_{t-1}^c \right),
\end{equation}
where $p$ and $h$ represent the densities of the hidden and observed data, respectively.
The next section proposes special cases of (\ref{Modgene}) appropriate for animal movement data.

\subsection{A General Directional Random Walk Model}
\label{sec:Obsprocess}

In this section, we present some new angular-distance specifications for the joint density of (\ref{Modgene}). The proposal relies on the following assumptions:

\begin{itemize}
\item[(A1).] Given the hidden process $S_{0:T}$, the observed processes $y_{0:T}$ and $d_{0:T}$ are independent, i.e.,
\begin{equation}\label{hfg}
h(y_t,d_t|S_t,\mathcal{F}_{t-1}^c)=f(y_t|S_t,\mathcal{F}_{t-1}^c)g(d_t|S_t,\mathcal{F}_{t-1}^c), t=1,\ldots,T.
\end{equation}
Moreover we suppose that the observed processes $y_{0:T}$ and $d_{0:T}$ are Markovian of order 1 with respect to the hidden process $S_t$,$t=1,\ldots,T$:
\item[(A2).] Given the hidden process $S_{0:T}$, we suppose that
\begin{equation}\label{hfg}
f(y_t|S_t,\mathcal{F}_{t-1}^c)=f(y_t|S_t,\mathcal{F}_{t-1}^o), t=1,\ldots,T.
\end{equation}
The Markovian assumption of order one is given by the distribution of $y_t$ only depend on the present hidden state $S_t$ and not $S_{t-1}, \ldots, S_0$.

\item[(A3).] Given the hidden process $S_{0:T}$, we suppose that
$$
g(d_t|S_t,\mathcal{F}_{t-1}^o)=g(d_t|S_t), t=1,\ldots,T.
$$
\end{itemize}
By assumption (A3), the distance $d_t$ is independent of $d_{t-s}$, $s = 1,\ldots, t$  for every time step $t=1,\ldots,T$. We made this assumption for computational time to be reasonable, otherwise we would have to model the distances as an autoregressive process. Let $g_k$ denote the density of $d_t$ given that the
hidden process is in state $k$ at time $t$. 
For the observed directions $y_{0:T}$, according to Assumption (A2), $f(y_t|S_t,\mathcal{F}_{t-1}^o)$ depends on $S_t$ but also on $\lbrace y_{t-s} \rbrace_{s<t}$ and on environmental variables observed in $\mathcal{F}_{t-1}^o$. Let $f_k(.|\mathcal{F}_{t-1}^o)$ be the density of  ${y}_t$ given the information in $\mathcal{F}_{t-1}^o$ knowing that the hidden process is in state $k$
at time $t$.

%
We now propose specific parametric forms for the functions $f_k$ and $g_k$. For $g_k$, any density
function on the positive real line can be used. We use (as \cite{Langrock2012}) Weibull  and gamma distributions in the data analysis section, while we use an exponential distribution for the simulation study because estimation of its parameter is faster. We denote by $\lambda^{(k)}$ the parameters of the density $g_k$. For instance in the application of Section 5, $\lambda^{(k)}=(\mathbf{\lambda}_1^{(k)},\mathbf{\lambda}_2^{(k)})$ where $\lambda_1^{(k)}$ and $\lambda_2^{(k)}$ denote respectively the shape and the scale parameters of a gamma distribution. The construction of the conditional circular densities is discussed next.

\noindent { \bf Circular multivariate regression model}
\\
Circular regression models for BCRW (\cite{Langrock2012} Appendix C, \cite{Duchesne2015}) express $y_t$ as a von Mises distributed with mean direction depending on $y_{t-1}$ and on other explanatory angles plus a homogenous error whose distribution depends on a fixed concentration parameter $\kappa$.  \cite{Rivest2015} show that the log-likelihood for estimating the parameters of such models is often multi-modal, making parameter estimation problematic. Multimodal log-likelihoods also occur in multi-state models when the errors are assumed to be homogenous.  This is illustrated in \ref{a:C4}.  The solution to avoid these multimodal log-likelihoods is to adopt a consensus error model as defined in \cite{Rivest2015}. Knowing that the animal is in state $k$,   a consensus error model for $y_t$ depends on the vector

\begin{equation}\label{V}
\mathbf{V}_t^{(k)}=\kappa_0^{(k)} \left(\begin{array}{c}
\cos(y_{t-1})\\
\sin(y_{t-1})
\end{array}  \right)+\sum_{i=1}^p \kappa_i^{(k)} z_{it} \left(\begin{array}{c}
\cos(x_{it})\\
\sin(x_{it})
\end{array} \ \right) , t=1,\ldots,T,
\end{equation}
where $\mathbf{\kappa}^{(k)}=(\kappa_0^{(k)},\dots,\kappa_p^{(k)})$ are unknown parameters depending on the state $k$. The mean direction, denoted $\mu_t^{(k)}$, of the von Mises distribution is the direction of $\mathbf{V}_t^{(k)}$. The parameters $\kappa_i^{(k)} \in \mathbb{R}$ quantify the influence of  target $i$ on the animal's direction. When all $\kappa_i^{(k)}=0$, $i=1,\ldots,p$, then $\mu_t^{(k)}=y_{t-1}$ and the animal tends to move in the direction of its previous step; the model then simplifies to the correlated random walk model (CRW).
Conversely, if target $i$ is highly attractive, then $\kappa_i^{(k)}$ is large. Similarly, a strongly negative value of $\kappa_i^{(k)}$
means that the target $i$ has a repulsive effect and the animal tends to move away from it. We can remark that the vector (\ref{V}) can depend on angular explanatory variables ($x_{it}$) or real variables ($z_{it}$) as explained at the beginning of Section 2.

In a consensus model, the concentration parameter, denoted $\ell_t^{(k)}$, of the von Mises distribution is the length of $\mathbf{V}_t^{(k)}$. Thus the concentration parameter depends on the level of agreement between the various
directional targets. If all targets and $y_{t-1}$ point in the same direction, then the concentration $\ell_t^{(k)}$ is large and the distribution of $y_t$ is concentrated around the mean direction  $\mu_t^{(k)}$.
Under these assumptions, we can write the density $f_k(y_t|\mathcal{F}_{t-1}^o)$ as
\begin{equation}\label{fk}
f_k({y}_t|\mathcal{F}_{t-1}^o;\kappa^{(k)})=\frac{1}{2 \pi I_0(\ell_t^{(k)})} \exp \left\lbrace \ell_t^{(k)} \cos (y_t-\mu_t^{(k)})\right\rbrace, t=1,\ldots,T,
\end{equation}
where $I_0$ is the modified Bessel function of order 1 (\cite{Mardia00}, p. 36).
As shown by \citep{Rivest2015} and \citep{Duchesne2015}, since $\mu_t^{(k)}$ and $\ell_t^{(k)}$ are the direction and length of the same vector (\ref{V}) and (\ref{fk}) give
$$
f_k({y}_t|\mathcal{F}_{t-1}^o;\kappa^{(k)})=\frac{1}{2 \pi I_0(\ell_t^{(k)})} \exp \left\lbrace \kappa_0^{(k)} \cos (y_t-y_{t-1})+\sum_{i=1}^p  \kappa_i^{(k)} \ z_{it}\cos(y_t-x_{it}) \right\rbrace,
$$
for $t=1,\ldots,T$.
This parametrization yields a numerically stable model, as the $\kappa_j^{(k)}$ are now the canonical parameters
of a distribution in the exponential family.

We recover many models proposed in the literature as special cases of this von Mises consensus model. A  special case with a single state ($p \left( S_t | \mathcal{F}_{t-1}^c \right) =1$ in (\ref{Modgene})) is for example  Breckling (1989), who proposed an autoregressive consensus von Mises distribution where $x_{it}=y_{t-i-1}$, $i=0,1,\ldots,p-1$. Another example is the biased correlated random walk model (BCWR, see for example \cite{Turchin}). It has $h \left(y_t,d_t|S_t , \mathcal{F}_{t-1}^c \right)=f(y_t|y_{t-1},x_{1t})$, see also \cite{Duchesne2015} for additional discussion of the consensus model when $p=1$.

In a multistate framework, the covariate free model with $\mu_t^{(k)}=\mu^{(k)}$ and $\ell_t^{(k)}=\kappa^{(k)}$ is the HMM studied by
\cite{Holzmann2006}. 
The HMM and hidden semi-Markov model (HSMM) considered
by \cite{Langrock2012} are also special cases of (\ref{Modgene}) where
$p \left( S_t | \mathcal{F}_{t-1}^c \right) =p \left( S_t | S_{t-1} \right)$ and
$h\left(y_t,d_t|S_t , \mathcal{F}_{t-1}^c \right)=f(y_t|y_{t-1},x_0,S_t)g(d_t|S_t)$, with $x_0$ a bias towards a fixed location.
Models such as (\ref{Modgene}) apply beyond animal movement. \cite{ailliot2012} use (\ref{Modgene}) to model
a wind time series, where $y_t$ is the wind direction and $d_t$ is the wind speed. They have, $p \left( S_t | \mathcal{F}_{t-1}^c \right) =p \left( S_t | S_{t-1},y_{t-1} \right) $
and $h\left(y_t,d_t|S_t , \mathcal{F}_{t-1}^c \right)=f(y_t|S_t,\mathcal{F}_{t-1}^o)$. 
\subsection{Markov and Semi-Markov Processes for the Hidden States}
\label{ss:Markov}
The model that we propose for
the hidden process $S_{0:T}$ is a homogeneous Markov chain. At any time step $t$, the animal is in one of $K$ possible states $\{1,\ldots,K\}$ and $p(S_t|\mathcal{F}_{t-1}^c)=p(S_t|S_{t-1})$. It is convenient to describe $S_{0:T}$ as a homogeneous multinomial process in discrete time. Let us define the sequence $\{\mathbf{S}_t,t=0,\dots,T\}$ of multinomial vectors $\mathbf{S}_t=(S_{1t},\dots,S_{Kt})$ where we set $S_{jt}=1$ and $S_{j't}=0$ for all $j'\neq j$ when $S_t=j$, $j,j'=1,\ldots,K$. At time $t=0$, we set $\mathbb{P}(S_{k 0}=1)=(\pi_0)_k $, such that $(\pi_0)_k \geq 0$, $k=1,\ldots,K$ and $\sum_{k=1}^K (\pi_0)_k =1 $. For the rest of the development we suppose that the initial distribution of the hidden process $\lbrace (\pi_0)_k ,k=1,\ldots,K\rbrace$ is known. We also introduce the
transition probabilities $\pi_{hk}=\mathbb{P}(S_{k t}=1 | S_{h,t-1}=1)$, for $h,k=1,\dots,K$. The contribution of the hidden process to the complete data density
given by (\ref{Modgene}) is a function of $S_{0:T}$ and of the transition probabilities:
\begin{equation}\label{contribhidden}
\prod_{t=1}^Tp(S_t|\mathcal{F}^c_{t-1})= \prod_{t=1}^T \prod_{h=1}^K \prod_{k=1}^K \pi_{hk}^{S_{h,t-1}S_{kt}}.
\end{equation}
Although the methodology presented in this paper works for any number of states $K$, we focus on $K=2$ in the simulation and data analysis sections. In the case $K=2$, we use the notation introduced below: $\pi_{11}=1-q_1$ and $\pi_{22}=1-q_2$.

\noindent {\bf A semi-Markov extension}.
\\
In some applications (for example, the study of the movement of bison, \cite{Langrock2012}), a semi-Markov hidden process is more realistic.
In a semi-Markov model, $p(S_t|\mathcal{F}_{t-1}^c)=p(S_t|S_{t-1},\tau_{t-1})$, where $\tau_{t-1}$ is the animal's dwell-time in the
state that it occupies at time step $t-1$, i.e., the number of consecutive time steps spent in that state.

Following \cite{Langrock2011}, any semi-Markov process can be approximated
to a high degree of accuracy by a Markov process with an enlarged set of states.
Each state of the approximating Markov process corresponds to a pair $(S,Q)$ where $S=1,\dots,K$ is the state of the animal and $Q=1,\dots,m$
 is the number of time points since the animal has arrived in this state.
 In the numerical example section we consider two states, $S=1,2$, and assume that the two dwell time distributions are shifted negative binomial distributions with parameters that depend on $S$. The transition probabilities are then denoted  $\pi_{(g,k)(h,\ell)}(n_g,q_{g})$, $g,h=1,2$, where $n_h,q_{h}$ denote the parameters (size and probability, respectively) of the dwell time distribution in state $h$. The quantity $\pi_{(g,k)(h,\ell)}(n_g,q_{g})$ denotes the probability that the animal has been in the state $g$ for $k$ consecutive time steps and go into the state $h$ for $\ell$ consecutive time steps. If $n_h=1$ for $h=1,2$, then the semi-Markov process reduces to a Markov process.  Given that one is in state $(g,k)$, the probability of staying in state $g$ is then $ \pi_{(g,k)(g,k+1)}(1,q_{g})=\pi_{gg}(1,q_{g})=1-q_{g}$. Details of this approximation are given in the \ref{a:B}. 

\subsection{Global Model Properties}

\begin{figure}[H]
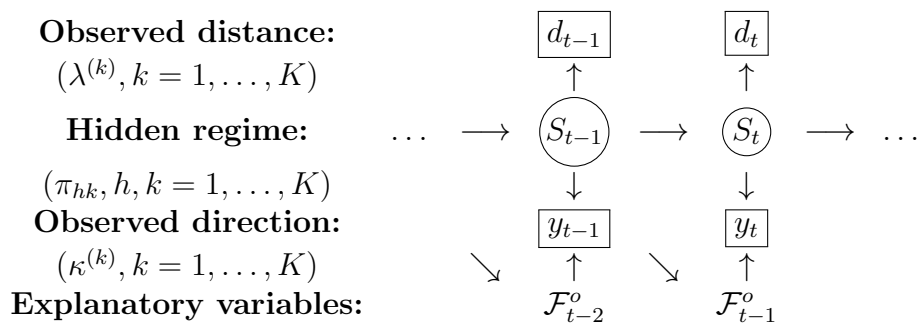

$$
\begin{array}{ccccccccc}
\textbf{Observed distance:} &  &  & \boxed{d_{t-1}} &  & \boxed{d_t} &  &   \\
(\lambda^{(k)}, k=1,\ldots,K)&  &  & \uparrow & & {\uparrow}&  &   \\
\textbf{Hidden regime:} & \dots & \longrightarrow & \encircle{$S_{t-1}$} & \longrightarrow & \encircle{$S_{t	}$} & \longrightarrow & \dots  \\
(\pi_{hk}, h,k=1,\ldots,K )& & & \downarrow & & {\downarrow}&  &  \\
\textbf{Observed direction:} &  &  &   \boxed{y_{t-1}} &  & \boxed{y_t} &  &

  \\
(\kappa^{(k)}, k=1,\ldots,K)&  & \searrow & \uparrow & \searrow& {\uparrow}&  &   \\
\textbf{Explanatory variables:} &  &  &   {\mathcal{F}_{t-2}^o} &  & {\mathcal{F}_{t-1}^o} &  &
\end{array}
$$
\caption{Dependence structure of the proposed model}\label{f:fig1}
\end{figure}

Figure~\ref{f:fig1} summarizes the dependence structure between the hidden process, the explanatory variables and the observed bivariate process
defined by (\ref{Modgene}).
We note that, in Figure~\ref{f:fig1}, given $S_t$ and $\mathcal{F}_{t-1}^o $, $S_{t-1}$ is independent of the observed data from time $t$ to $T$ (i.e. $\lbrace \mathcal{F}_{t+s}^o \rbrace_{s \geq 0} \setminus \mathcal{F}_{t-1}^o$). This is used in the implementation of the filtering-smoothing algorithm presented in the \ref{a:A}.

The proposed model is identifiable, in the sense that
(up to label switching) different values of the parameters will result in different joint distributions for the observed data. This is proved using an argument similar to that presented in  \cite{ailliot2012}.
We can also demonstrate that under the assumption that the hidden process $S_{0:T}$ is ergodic and there is only one fixed explanatory variable $\{(x_t,z_t)=(x_0,z_0),\ t=1,\ldots,T\}$ in the directional specification, then the observed data process
$(y_{0:T},d_{0:T})$ is also ergodic; through this condition is sufficient, simulations suggest that it is presumanly not necessary. In other words, in the long run (as $T$ gets large), the process converges to a stationary distribution.
In the case that the angular process $y_t$ is only predicted by past directions $y_{t-s}$, $s=1,\ldots,t$, then the consistency of the maximum likelihood estimators of the model parameters (up to label switching) is a simple application of Theorem 1 of \cite{Rydn2004}, who show the consistency of the MLE for auto-regressive processes.
\section{Inferential Procedures}
\label{s:inf}
We now propose a procedure to estimate $\mathbf{\theta}=(\mathbf{\pi},\mathbf{\kappa},\mathbf{\lambda})$, where $\mathbf{\pi}$ is a vector of $K \times (K-1)$ transition probabilities, $\mathbf{\kappa}$ are the $K \times (p+1)$ unknown parameters for the angular model and $\mathbf{\lambda}$ are the $2K$ parameters (scale and shape) of the model for the distance variable.  Given a series of observations
$\left\lbrace \left(y_t,d_t,\mathbf{x}_t,\mathbf{z}_t\right),t=0,\dots,T \right\rbrace$,
we can write the likelihood function of these parameters as a product of the one-step ahead predictive densities, see for instance \cite{hamilton2008regime}, which developed these models called ``regime switching models'' for the applications in econometrics,

\begin{equation}\label{Lobs}
\text{L}(\mathbf{\theta})=\prod_{t=1}^T \left( \sum_{k=1}^K f_k(y_t|\mathcal{F}_{t-1}^o,\kappa^{(k)})g_k(d_t,\lambda^{(k)}) \mathbb{P}(S_{tk}=1|\mathcal{F}_{t-1}^o,\mathbf{\theta}) \right).
\end{equation}
Where the ``predictive'' probabilities: $\mathbb{P}(S_{tk}=1|\mathcal{F}_{t-1},\theta)$, $t=1,\ldots,T$ are evaluated recursively using \ref{a:Predictive} in the filtering-smoothing algorithm given on \ref{a:A}, allowing the evaluation of (\ref{Lobs}).
The method used by \cite{Langrock2012} to evaluate the likelihood of their multi-state model and to estimate its parameters can be generalized to find the parameter values that maximize (\ref{Lobs}).  However, this method  sums the unobserved states out of the likelihood and does not allow to predict, at each time point, the underlying state of the animal  which has interesting ecological applications.  This is, however, possible with the EM algorithm hence our use of EM to maximize (\ref{Lobs}).

\subsection{EM Algorithm}
The EM algorithm is generally used for the maximization of likelihood functions when some data are missing or unobserved. Here, $y_{0:T}$ and $d_{0:T}$ are the observed
data and $S_{0:T}$ is the missing data.
The EM algorithm only requires evaluation of the complete data log-likelihood function, which in our case is easily derived from (\ref{Modgene}):
\begin{eqnarray*}
\log \text{L}_{\text{complete}}(\mathbf{\theta};\mathcal{F}_T^c) &=&  \sum_{t=1}^T \sum_{h=1}^K \sum_{k=1}^K S_{h,t-1}S_{k,t} \log \pi_{hk}(n_h,q_{h}) \\
&+& \sum_{t=1}^T \sum_{k=1}^K S_{kt} \log f_k({y}_t|\mathcal{F}_{t-1}^{o},\mathbf{\kappa}^{(k)})  \\
&+& \sum_{t=0}^T \sum_{k=1}^K S_{kt} \log g_k({d}_t|\mathbf{\lambda}^{(k)}).
\end{eqnarray*}
The EM algorithm consists of iterating an expectation (E) and a maximization (M) step. Let us denote by
$\hat{\mathbf{\theta}}_s$ the value of the estimate of $\mathbf{\theta}$ after the $s$-th iteration of the algorithm. Then the $(s+1)$-th iteration of the algorithm starts with one application of the E-step, which evaluates the expectation of $\log \text{L}_{\text{complete}}$ with respect to the conditional distribution of the missing values given the observed data, as follows:
\begin{eqnarray*}
Q(\mathbf{\theta}|\hat{\mathbf{\theta}}_s )&=& \mathbb{E}_{S_{0:T}}\left[\log \text{L}_{\text{complete}}(\mathbf{\theta};\mathcal{F}_T^c)|\mathcal{F}_{T}^o,\hat{\mathbf{\theta}}_s \right] \nonumber \\
&=& \sum_{t=1}^T \sum_{h=1}^K \sum_{k=1}^K \mathbb{E} (S_{h,t-1}S_{k,t}|\mathcal{F}_{T}^o,\hat{\mathbf{\theta}}_s) \log \pi_{hk}(n_h,q_{h}) \\
&+& \sum_{t=1}^T \sum_{k=1}^K \mathbb{E}( S_{kt}|\mathcal{F}_{T}^o,\hat{\mathbf{\theta}}_s) \log f_k({y}_t|\mathcal{F}_{t-1}^o,\mathbf{\kappa}^{(k)})  \\
&+& \sum_{t=0}^T \sum_{k=1}^K \mathbb{E}( S_{kt}|\mathcal{F}_{T}^o,\hat{\mathbf{\theta}}_s) \log g_k({d}_t|\mathbf{\lambda}^{(k)}).
\end{eqnarray*}
Then the value of $\hat{\mathbf{\theta}}_{s+1}$ is calculated in the M-step as the value of $\mathbf{\theta}$ that maximizes
 $Q(\mathbf{\theta}|\hat{\mathbf{\theta}}_s )$.

\subsubsection{• E step}
The function $Q(.|\hat{\mathbf{\theta}}_s)$ involves two conditional expectations, $\mathbb{E}( S_{kt}|\mathcal{F}_{T}^o,\hat{\mathbf{\theta}}_s) $ and $\mathbb{E} (S_{h,t-1}S_{k,t}|\mathcal{F}_{T}^o,\hat{\mathbf{\theta}}_s)  $. These can be efficiently computed by a forward-backward (filtering-smoothing) algorithm for Markov chains, see \cite{Sylvia13}. The filtering-smoothing algorithm starts from the initial time $t=0$ and computes the ``filtering" probabilities $\mathbb{P}(S_t|\mathcal{F}_t^o)$ by using predictive probabilities $\mathbb{P}(S_t|\mathcal{F}_{t-1}^o)$ (going forward in time). The last filtering probability $\mathbb{P}(S_T|\mathcal{F}_T^o)$ is then used to compute the ``smoothing" probabilities $\mathbb{P}(S_t|\mathcal{F}_T^o)$ using Bayes theorem (going backward in time).
We outline the details of this implementation of the E-step in the  \ref{a:A}.

\subsubsection{• M step}
 For the M-step, we see that $Q(\mathbf{\theta}|\hat{\mathbf{\theta}}_s ) $ is a sum of three functions that depend on different sets of parameters and can thus be maximized separately:

\begin{itemize}

\item When the latent states follow a Markov process, there is a closed form expression for the maximizer of the hidden process part,
\begin{equation}
\label{pihat}
\hat{\pi}_{hk}^{(s+1)}=\frac{\sum_{t=1}^T  \mathbb{E} (S_{h,t-1}S_{k,t}|\mathcal{F}_{T}^o,\hat{\mathbf{\theta}}_s)}{\sum_{t=1}^T \mathbb{E} (S_{h,t-1}|\mathcal{F}_{T}^o,\hat{\mathbf{\theta}}_s)}, h,k=1,\ldots,K,
\end{equation}
which represent the expected number of transitions from state $h$ to state $k$ divided by the expected number of transitions leaving from state $h$.

\item Since $f_k({y}_t|\mathcal{F}_{t-1}^o)$ has a consensus von Mises density, the log-likelihood for the directional part is concave and the maximum is easily calculated; details are available in \cite{Rivest2015}.

\item The maximization algorithm depends on $g_k(d_t|\mathbf{\lambda}^{(k)})$. For the exponential distribution used in the simulation study, the maximizer has a closed form expression. For the Weibull or gamma distributions used in the data analysis, a numerical maximization (e.g., Newton-Raphson algorithm) of the weigthed log-likelihood is needed to compute the estimates.

\end{itemize}

\textbf{Semi-Markov specification}

If the hidden process model is not saturated (e.g., semi-Markov specification), then $\hat{n}_h^{(s+1)},\hat{q}_{h}^{(s+1)}$ cannot be computed explicitly from the function $Q(.|\hat{\mathbf{\theta}}_s ) $
and numerical maximization is usually needed.
\subsection{Sampling Distributions}

Quantities that are usually required for inference such as the value of the maximized log-likelihood for the observed data or an estimation of the variance matrix of $\hat{\mathbf{\theta}}$ are not
directly computed when using the EM-algorithm. 
The filtering-smoothing algorithm is used to evaluate the likelihood for the observed data (\ref{Lobs}). Moreover at each time $t$, one can evaluate the probability that the animal is in state $k$ using the value of $\mathbb{E}(S_{kt}|\mathcal{F}_T)$ in the ``smooth'' part of the algorithm (see \ref{a:smooth}).
 Because we are able to compute $\log \text{L}(\hat{\mathbf{\theta}}_{\text{MLE}})$, we can numerically approximate the negative of its Hessian matrix,
whose inverse, denoted ${v}$, is the usual estimate of the variance matrix of the maximum likelihood estimators. A numerical approximation of the Hessian matrix is available under most software implementations
of the Broyden--Fletcher--Goldfarb--Shanno (BFGS) algorithm  (\cite{avriel2003nonlinear});  in the data analysis section
 we use the one provided in the R function \texttt{optim}.
 
\subsection{Model Selection}

The number of hidden states is usually unknown and its determination is a difficult problem. We follow many applications (\cite{Langrock2012}, \cite{Holzmann2006}, \cite{ailliot2012}) to animal or wind movement and assume two states for the hidden process. In animal movement studies, a two state model is stable and interpretable.
The selection of the  potential directional targets can be done using the classical criteria (AIC, BIC) or Wald's tests.
\section{Simulation Study}
\label{s:simul}

This section reports the results of a simulation study that investigates the sampling properties of estimators presented in Section \ref{s:inf}. We simulated
the movement of one animal in the plane. The ``target'' was placed at the center of a map and the covariate $x_t$ represents the direction to this target
at time step $t$. Then each simulation scenario consisted in repeating the following steps 500 times: (i) a two-state Markov chain $S_{0:T}$ with transition matrix $\mathbf{\pi}$ is generated; (ii) at time 0, the animal is placed at a random position close to the south west corner of the map; (iii)
at each time step $t$, $t=1,2,\ldots$, the location of the animal is obtained by simulating a direction $y_t$ and a distance $d_t$ from the Markov switching model proposed in Section \ref{sec:Obsprocess} with
$y_t$ generated according to a consensus von Mises model with parameters $\kappa_0^{(k)},\kappa_1^{(k)}$ and explanatory angles $y_{t-1}$ and $x_t$ and $d_t$ is simulated from an exponential distribution with mean $\lambda^{(k)}$ units; (iv) the simulation stops when the animal is within 30 distance units from the target.

We considered two different simulation scenarios. The values of the parameters used in each one are given in Table~\ref{t:scenario}.
Scenario 1 is one where the animal shows high directional persistence and high attraction to the target when in state 1, and high directional persistence and a moderate repulsion from the target in state 2. The second scenario shows moderate directional persistence and moderate attraction to the target when in state 1, and little directional persistence and a weak attraction toward the target in state 2. A characteristic trajectory of one simulation under the first scenario is presented in \ref{a:C5} since this scenario present attractive and repulsive effect of the target on the trajectory of the animal in each state.

\begin{table}[H]
\caption{Parameters for the two simulation scenarios.}
\centering
\begin{tabular}{c c c c  }
\hline

\hline
 & \textbf{parameter}& \textbf{scenario 1} & \textbf{scenario 2}   \\
\hline
& & & \\
 &$P= \left( \begin{array}{cc}
 1-q_{1} & q_{1} \\
 q_{2} & 1-q_{2}
 \end{array} \right)$  & $\left( \begin{array}{cc}
 0.9 & 0.1 \\
 0.2 & 0.8
 \end{array} \right)$ & $\left( \begin{array}{cc}
 0.6 & 0.4 \\
 0.1 & 0.9
 \end{array} \right)$   \\
 &  &  &    \\
 & $\kappa_0^{(1)}$ & 20 & 5  \\
 & $\kappa_1^{(1)}$ & 10 & 4.5  \\
&$\lambda^{(1)}$ &0.7  &2  \\
 &  &  &  \\
&$\kappa_0^{(2)}$ & 15 & 2  \\
  &$\kappa_1^{(2)}$ & -6.5 & 0.4   \\
&$\lambda^{(2)}$ & 1.2 & 5  \\
\hline
\hline
\end{tabular}
\label{t:scenario}
\end{table}

The average length of the series is 533 under the first scenario and 695 under the second scenario, the minimum  (resp. maximum) was 420 (resp. 630) steps in the first scenario, for the second scenario the minimum was 580 (resp. 870) steps. The model was fitted to each simulation sample and the following statistical indicators were calculated:
\begin{eqnarray}
b(\hat{\theta})&=& \frac{1}{500} \sum_{i=1}^{500}(\hat{\theta}^{(i)}-\theta) \label{biais}\\
\text{Sd}(\hat{\theta})&=& \sqrt{\frac{1}{499} \sum_{i=1}^{500}(\hat{\theta}^{(i)}-\bar{\theta})^2 } \label{RMSE}\\
\hat{\mathbb{E}}\left[\sqrt{v(\hat{\theta})}\right]&=& \frac{1}{500} \sum_{i=1}^{500} \sqrt{v(\hat{\theta}^{(i)})} \label{sdapprox},
\end{eqnarray}
with $\hat{\theta}^{(i)}$ the parameter estimate in the $i$-th simulation and $\bar{\theta}$ the mean of the estimates over the 500 simulations. Equation (\ref{biais}) gives the bias of the estimator, (\ref{RMSE}) its standard deviation and (\ref{sdapprox}) is the mean of the standard error estimator. We also computed the coverage of the nominal 95\% Wald confidence interval as the proportion of times that the true parameter $\theta$  belonged to  $\hat \theta^{(i)}\pm 1.96 v(\hat{\theta}^{(i)})$. The results are summarized in Table \ref{t:paramscenario}.

\begin{small}
\begin{table}[H]
\caption{Result of the $N=500$ simulations on the two simulation scenarios.}

\centering
\begin{tabular}{c c c c c c c c }
\hline

\hline
 & & $\theta$ & $b(\hat{\theta})$ & $\text{Sd}(\hat{\theta})$ & $\mathbb{E}\left[\sqrt{v(\hat{\theta})}\right]$  & $\begin{array}{c}
 95\% \text{interval}\\
  \text{coverage}
 \end{array}  $  \\
\hline
\hline
 \multirow{8}{*}{ $\begin{array}{c}
 \textbf{scenario 1}  \\
  (\bar{n}=533)
 \end{array}  $  } &$q_{1}$ & 0.1 & 0.002  & 0.020 & 0.021 & 0.950   \\
 &$q_{2}$ & 0.2 &-0.003  & 0.039 &0.039 &  0.954 \\

 & $\kappa_0^{(1)}$ & 20 & 0.259 & 1.675 & 1.718 & 0.958 \\
 & $\kappa_1^{(1)}$ & 10 & 0.125 & 1.089 & 1.051 &  0.960 \\
&$\lambda^{(1)}$ &0.7  & -0.002  & 0.038 &  0.037 &0.950  \\
 & $\kappa_0^{(2)}$ & 15 & 0.322 &2.065  & 1.944 & 0.950  \\
 & $\kappa_1^{(2)}$ & -6.5 & -0.098 &1.158 & 1.113 & 0.940  \\
&$\lambda^{(2)}$ & 1.2 & 0.014 &0.110 &0.111 & 0.952  \\
\hline
\hline
 \multirow{8}{*}{ $\begin{array}{c}
 \textbf{scenario 2}  \\
  (\bar{n}=695)
 \end{array}  $ } &$q_{1}$ & 0.4 & -0.008  & 0.060  & 0.058&  0.950 \\
 &$q_{2}$ & 0.1 & -0.002 &  0.020 & 0.022& 0.958  \\
 & $\kappa_0^{(1)}$ &5  & 0.137 & 0.864 & 0.804 & 0.950 \\
 & $\kappa_1^{(1)}$ & 4.5 &0.136  & 0.888 & 0.853 & 0.940  \\
&$\lambda^{(1)}$ & 2 & -0.003 & 0.162 & 0.167 & 0.956 \\
 & $\kappa_0^{(2)}$ & 2 &0.003  &  0.071 & 0.075 &  0.966 \\
 & $\kappa_1^{(2)}$ & 0.4 & -0.007 & 0.068 &  0.072 &0.958   \\
&$\lambda^{(2)}$ & 5 & 0.010 & 0.185&0.186 &0.952  \\
\hline
\hline
\end{tabular}
\label{t:paramscenario}
\end{table}
\end{small}

Table~\ref{t:paramscenario} shows that the inferential procedure presented has good statistical properties. First the maximum likelihood estimator appears to be unbiased ($b(\hat{\theta}) \approx 0 $)  for the coefficient of distance and for the transition probabilities. Estimates of the $\kappa$ parameters exhibit a weak bias, which is higher in the first scenario, a phenomenon well known for the von Mises distribution (\cite{Mardia00}). The quasi equality between Sd$(\hat{\theta})$ and $\mathbb{E} \left[\sqrt{v(\hat{\theta}}) \right]$ indicates that our variance matrix approximation of the variance of the parameter estimates is correct. Finally, the 95\% Wald confidence intervals based on $v$ have an empirical coverage rate that is approximately equal to 95\%.

\section{Application to the Analysis of the Movement of Caribou}
\label{s:caribou}
\begin{figure}[H]
\centering
\includegraphics[scale=0.4]{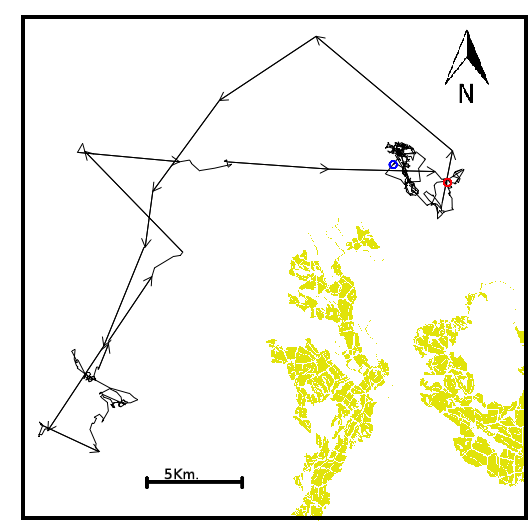}
\caption{Caribou trajectory from December 28 2005 to April 15 2006. The red circle corresponds to the start and the blue one to the end of the observed trajectory. The yellow color corresponds to the locations with regenerating cuts.}
\label{f:trajobs}
\end{figure}

We now apply the model  using data collected by \cite{Latombe2014} about the movement of forest caribou in the Cote-Nord region of Quebec, Canada. This involved 23 animals with different home ranges.  The proposed two state model fitted well the data of several animals.  We observed strong variation among animals so in this section we focus on a single animal, wearing a collar recording its locations every four hours, observed in the 2006-7 winter period. Its position was observed at $T=617$ time points and Figure \ref{f:trajobs} shows its trajectory, with distance expressed in meters, during the observation period. In Figure~\ref{f:trajobs} the distances traveled in four hour intervals are generally small and the caribou is mostly observed around two different sites.  Because the individual travels between sites, two different states, encamped and traveling, might be envisaged.  The animal movement mostly takes place in the NE-SW direction.

Besides directional persistence, $y_{t-1}$, several explanatory angles were considered in the analysis.  The ones kept for the final model were $x_{\text{cut}}$, the direction to the closest regenerating cut (i.e., a forest stand that has been cut between 5 and 20 years ago, the yellow in Figure~\ref{f:trajobs}), and the direction to the centroid of a cluster of recently visited locations $x_{\text{center} }$. 
At time $t$ the locations visited by the animal between times 0 and $t-1$ are put in clusters.  Cluster 1 is the set of locations visited between time 0 and $t_1$.  The cluster ends at time $t_1$ means that the distance between the position at time $t_1+1$ and the centroid of the cluster is, for the first time, larger than a fixed number $D=1.6$ km (see the Appendix of \cite{Latombe2014} for more details).  In the same way the second cluster is made of the locations visited between time $t_1+1$ and $t_2$ and so on.  At time $t$, the average locations for all clusters are calculated and the cluster whose average location is closest to the current animal position is used to compute  $x_{\text{center} }$ as the direction to the average location of the closest cluster.
%
%
 The exploratory analysis in the \ref{a:D} reveals that the relative appeal for all these targets varies with the distance traveled.  This suggests fitting a two state model to the data.
 \subsection{Analysis with a Two-State Model}

We now fit the proposed model with $K=2$ states featuring  directional persistence and the two explanatory angles, $x_{\text{cut}}$ and $x_{\text{center} }$, presented above. The distance traveled was fitted using the Weibull and gamma distributions, both AIC and BIC criteria select the gamma distribution (AIC$_{\text{Weibull}}=1620.60$ and BIC$_{\text{Weibull}}=1673.70$). Two models, with a Markov and a semi-Markov specification for the hidden process, are considered.
The results are summarized in Table~\ref{t:resultCar}.

\begin{table}[H]
\caption{Estimation of the parameters of multistate models, with gamma distributed distances; $\mathbf{\lambda}_1^{(k)}$ and $\mathbf{\lambda}_2^{(k)}$ denote the shape and scale parameters of the gamma distribution in the state $k$; $n_k$ and $q_k$ are the size and probability of the negative binomial distribution of the state $k$'s dwell time. }
\centering
\begin{tabular}{c c  c  c c }
\hline

\hline
 & \multicolumn{2}{c}{Markov}  & \multicolumn{2}{c}{Semi-Markov} \\
\cline{2-5}
 & Estimate & S.e. & Estimate & S.e. \\
\hline
 $q_{1}$ & 0.2799 & 0.0891& 0.1290& 0.1087 \\
  $n_{1}$ &1 & . &  0.3046& 0.2529 \\
$q_{2}$  & 0.0231  & 0.0097  & 0.0157 &0.0124\\
    $n_{2}$ & 1 &.  &  0.6034& 0.3828\\
     & & &  & \\
  $\kappa_{\text{persist.}}^{(1)}$ & 1.2666 & 0.3327  & 1.2617& 0.3368     \\
 $\kappa_{\text{center}}^{(1)}$ & 0.3732 & 0.2994 &  0.4251&  0.2995 \\
  $\kappa_{\text{cut}}^{(1)}$ & 0.1601 & 0.3013 & 0.1274 &  0.3043 \\
$\mathbf{\lambda}_1^{(1)}$& 0.6477 &0.1372 &  0.6670&  0.1366\\
$\mathbf{\lambda}_2^{(1)}$& 3.0444 &0.8450   & 2.9361& 0.8122 \\
     & & & & \\
$\kappa_{\text{persist.}}^{(2)}$ & 0.0274 & 0.0618 &0.0263 &0.0619\\
 $\kappa_{\text{center}}^{(2)}$  & 0.2590    &  0.0694&  0.2563  & 0.0695\\
  $\kappa_{\text{cut}}^{(2)}$ & 0.1454 & 0.0679 &   0.1462&0.0698 \\
$\mathbf{\lambda}_1^{(2)}$  &1.2626  &  0.0774& 1.2704&0.0779 \\
$\mathbf{\lambda}_2^{(2)}$  &0.1351  &  0.0119 & 0.1332&0.0119 \\
\hline
\multicolumn{1}{c}{($l$,AIC,BIC)} & \multicolumn{2}{c}{(-794.91,1613.82,\textbf{1666.91})} &   \multicolumn{2}{c}{(-793.11,1614.22,1676.17)}  \\
\hline
\hline
\end{tabular}
\label{t:resultCar}
\end{table}

\begin{figure}[H]
   \begin{minipage}[c]{.46\linewidth}
\includegraphics[scale=0.6]{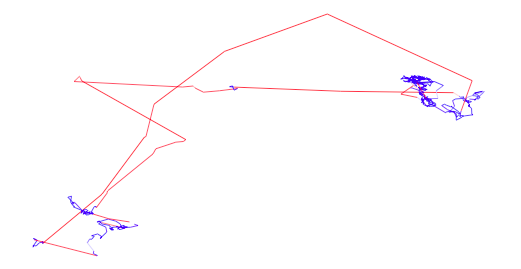}
   \end{minipage} \hfill
   \begin{minipage}[c]{.2\linewidth}
      \includegraphics[scale=0.5]{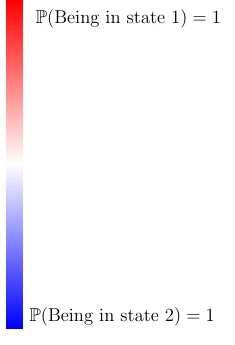}
   \end{minipage}
   \caption{Hidden state probabilities at each time step of the trajectory of the caribou.}
\label{f:comportementestim}
\end{figure}

In Table~\ref{t:resultCar}, state 1 is a traveling mode with a large estimate for the average distance traveled $\hat{\mathbf{\lambda}}_1^{(k)} \hat{\mathbf{\lambda}}_2^{(k)}$, while state 2 is encamped.  Most of the data points are in state 2 and the information for the estimation of the state 1 parameters is limited.  The final model for data interpretation, selected as the one with the smallest BIC, is the one with a Markovian hidden process. Table~\ref{t:resultCar} shows two different regimes for the direction and the traveled distance between steps. In the first state, the caribou has a strong significant directional persistence ($\hat{\kappa}_{\text{persist.}}^{(1)}>1.26, \ s.e.=0.33 $). The two $\kappa$ parameters for the environmental targets are positive; however, because of the limited data available in state 1, their standard errors are large and these parameters are not significantly different from 0. In state 1, the animal moves at an average speed of  $\hat{\mathbf{\lambda}}_1^{(1)} \hat{\mathbf{\lambda}}_2^{(1)} /4\approx 493 $ m per hour.
\\
\\
In the second regime the animal is almost stationary, moving by about 43 m ($\hat{\mathbf{\lambda}}_1^{(2)} \hat{\mathbf{\lambda}}_2^{(2)}/4$) per hour.  The directional persistence parameter ${\kappa}_{\text{persist.}}^{(2)}$ is not significantly different from zero. The parameters for the two environmental targets are significantly larger than 0, suggesting that the caribou is attracted by locations previously visited and by the closest regenerating cut. The interpretation of the significant parameter for wood regenerating cuts is challenging as the animal never reaches them (Figure \ref{f:trajobs}).  It could be an accidental relationship:  when in the encamped state in the north-east corner of the map, the caribou moves in the north-south direction and the wood regenerating cuts happen to be located south of this area.  Considering \cite{Fortin13}, the wood regenerating cuts could also act as a proxy for the center of the animal’s current home range, the area where the caribou relocated after the cuts.
\\
\\
The stationary distribution of the latent fitted Markov chain is in state 1 with probability $\hat q_2/( \hat q_2+\hat q_1)=0.076$, showing that  the caribou was observed traveling about 47 out of $T=617$ sightings. This  explains why the parameter estimators have a low precision in state 1. The state of the localization at time $t$ can be identified using  the smooth probabilities $\mathbb{P}(S_{tk}=1 | \mathcal{F}_T; \hat{\mathbb{\theta}}_{\text{MLE}}) $, $t=1,\ldots,T$, $k=1,2$ calculated in the filtering-smoothing part of the E-step of the EM algorithm. This is depicted in Figure~\ref{f:comportementestim} with a color gradient from red (state 1, $ \mathbb{P}(S_{t1}=1 | \mathcal{F}_T; \hat{\mathbb{\theta}}_{\text{MLE}})=1$) to blue (state 2, $ \mathbb{P}(S_{t2}=1 | \mathcal{F}_T; \hat{\mathbb{\theta}}_{\text{MLE}})=1$).

\cite{Latombe2014} analyzed these data with single state models.  They found directional persistence to be an important determinant of caribou movement. Our analysis complements theirs as we show that this variable is important only when the caribou travels between sites (i.e., when the animal is in state 1).  For more than 90\% of the observation times, the caribou was in state 2 (encamped mode) and its movement was not determined by directional persistence. The angular analysis clearly identified two environmental features, regenerating cuts and previous locations visited, that influenced (attraction effect) caribou movement.
Plots produced by our method, such as the one shown in Figure \ref{f:comportementestim}, can thus be used to identify areas of inter-patch movements (state 1, red) or of residency
(state 2, blue) which tend to be studied with two different models in ecological research (e.g., \cite{BastilleRousseau2011},  \cite{Dancose2011} and \cite{Courbin2014}).

Section 4 shows that the proposed model describes well the motion of animal going towards a target.  The data analyzed in this section, see Figure 2, is about an animal going back and forth between two targets.  Is model (\ref{Lobs}) suitable for such data? \ref{a:valid} shows that it is,  through simulations.  If an animal is attracted by two targets and if the model at step $t$ shows a directional bias only towards the closest one, then (\ref{Lobs})  describes the motion of an animal moving back and forth between the two targets.  See  \ref{a:valid} for details.

\section{Conclusion}

This paper proposes a multi-state model with a general directional specification to describe the movement of an animal. It improves on classical BCRW because it allows the animal to exhibit several movement behaviors.   It is a general method to determine different behaviors, as it can handle and reveal the response of an animal to an arbitrary number of environmental features. The method generalizes the contribution of (\cite{Langrock2012}) whose statistical model, based on HMMs, only permits the inclusion of one environmental characteristic in the analysis.

In ecology, the understanding of the interplay between animal movement and habitat heterogeneity, including the characterization of strategies that animals use to locate sites for forage and safety, has been a long standing problem (\cite{smith}). By using recent techniques for the implementation of the EM algorithm in complex settings, we provide new statistical tools to identify these hidden animal behaviors. The appeal of these new methods is that they can handle an arbitrary number of explanatory variables associated to the animals' environment. These explanatory variables can be directions to environmental features, or continuous variables accounting for a time effect of some habitat characteristics.  See \cite{Rivest2015} for a more general discussion of single-state angular regression models featuring both angular and continuous explanatory variables.

The directional component of the proposed angular-linear random walk process has several important properties.   It is based on a sound statistical method to combine an arbitrary number of explanatory variables. From a numerical point of view, the M-component of the EM algorithm is simple as it involves the maximization of several convex functions, at least in the Markovian case. This is so because of the consensus error specification for the angular part of the model. A regression parameter $\kappa$ is easily interpreted as the relative appeal of a particular target when compared to the others.  The $\kappa$ estimates allow an accurate characterization of the different behavioral modes of an animal, as illustrated in the analysis of the caribou data.

\cite{barraquand2008animal} point out that because of the heterogeneity of the landscape, animals have to move through various types of areas that are more or less suitable for their current needs.  They propose a technique to identify intensively used areas based on distance traveled. The smooth probabilities obtained through the EM algorithm are an alternative to identify these intensively used places that takes into account both the trajectory and the landscape heterogeneity.
 
Technological advances in satellite telemetry, such as Argos archival data loggers, have allowed researchers to track animal movements and behavior in environments that are difficult to study like marine systems (\cite{PATTERSON2008}, \cite{Albertsen2015}). Measurement errors in the locations acquired with this technology should be taken into account (\cite{Jonsen2005},  \cite{Albertsen2015}). In our application, the land animal is followed using the more accurate GPS technology. In this case, the errors in the locations are small in comparison to the image resolution,  and this translates into errors in the directions and distances between locations and targets that are negligible.
 
There are several possibilities for further extension of the method presented here.
Because caribou exhibit heterogeneity in their movement behavior, the simultaneous analysis of the movement of many individuals would require a model with random effects. Including these in the proposed multi-state model should be relatively straightforward, but adapting the numerical procedure appears difficult. 

When a semi-Markov structure is assumed for this hidden process, the Markov approximation considered here involves a large number of states and is computationally demanding. Or when trying to model the behavior of an animal over a long period of time (e.g., more than one ``biological season'', see \cite{Basille2012}), the time homogeneity assumption
can be unreasonable. Hence defining a directional model based on a more complex hidden process could be an extension. Though the numerical algorithm proposed
here works really well when the hidden process is a time-homogeneous Markov chain, a new numerical approach would presumably be required if a different hidden
process were assumed.

\section*{Acknowledgements}

The authors are grateful to L\'ea Harvey for her help with the caribou data
and to Guillaume Latombe and Marie-Caroline Prima for insightful discussions.  Financial support was provided by a grant from the Fonds de recherche du Qu\'ebec - Nature et technologie to Louis-Paul Rivest, Thierry Duchesne and Daniel Fortin, and a scholarship from the Institut des sciences math\'ematiques awarded to Aur\'elien Nicosia. Field work was supported by the Natural Sciences and Engineering Research Council of Canada (NSERC) - Universit\'e Laval Industrial Research Chair in Boreal Forest Silviculture and Wildlife.

\bibliographystyle{unsrtnat}

\bibliography{mybib}

\begin{thebibliography}{32}
\providecommand{\natexlab}[1]{#1}
\providecommand{\url}[1]{\texttt{#1}}
\expandafter\ifx\csname urlstyle\endcsname\relax
  \providecommand{\doi}[1]{doi: #1}\else
  \providecommand{\doi}{doi: \begingroup \urlstyle{rm}\Url}\fi

\bibitem[Nathan et~al.(2008)Nathan, Getz, Revilla, Holyoak, Kadmon, Saltz, and
  Smouse]{Nathan2008}
R.~Nathan, W.~M. Getz, E.~Revilla, M.~Holyoak, R.~Kadmon, D.~Saltz, and P.~E.
  Smouse.
\newblock A movement ecology paradigm for unifying organismal movement
  research.
\newblock \emph{Proceedings of the National Academy of Sciences}, 105\penalty0
  (49):\penalty0 19052--19059, dec 2008.
\newblock \doi{10.1073/pnas.0800375105}.
\newblock URL \url{http://dx.doi.org/10.1073/pnas.0800375105}.

\bibitem[Latombe et~al.(2014)Latombe, Parrott, Basille, and
  Fortin]{Latombe2014}
Guillaume Latombe, Lael Parrott, Mathieu Basille, and Daniel Fortin.
\newblock Uniting statistical and individual-based approaches for animal
  movement modelling.
\newblock \emph{{PLoS} {ONE}}, 9\penalty0 (6):\penalty0 e99938, jun 2014.
\newblock \doi{10.1371/journal.pone.0099938}.
\newblock URL \url{http://dx.doi.org/10.1371/journal.pone.0099938}.

\bibitem[Holyoak et~al.(2008)Holyoak, Casagrandi, Nathan, Revilla, and
  Spiegel]{Holyoak2008}
M.~Holyoak, R.~Casagrandi, R.~Nathan, E.~Revilla, and O.~Spiegel.
\newblock Trends and missing parts in the study of movement ecology.
\newblock \emph{Proceedings of the National Academy of Sciences}, 105\penalty0
  (49):\penalty0 19060--19065, dec 2008.
\newblock \doi{10.1073/pnas.0800483105}.
\newblock URL \url{http://dx.doi.org/10.1073/pnas.0800483105}.

\bibitem[Turchin(1998)]{Turchin}
Peter Turchin.
\newblock \emph{Quantitative Analysis of Movement: Measuring and Modeling
  Population Redistribution in Animals and Plants}.
\newblock Beresta Books, 1998.
\newblock ISBN 0996139508.

\bibitem[Jonsen et~al.(2005)Jonsen, Flemming, and Myers]{Jonsen2005}
Ian~D. Jonsen, Joanna~Mills Flemming, and Ransom~A. Myers.
\newblock Robust state-space modeling of animal movement data.
\newblock \emph{Ecology}, 86\penalty0 (11):\penalty0 2874--2880, nov 2005.
\newblock \doi{10.1890/04-1852}.
\newblock URL \url{http://dx.doi.org/10.1890/04-1852}.

\bibitem[Shimatani et~al.(2012)Shimatani, Yoda, Katsumata, and
  Sato]{Shimatani2012}
Ichiro~Ken Shimatani, Ken Yoda, Nobuhiro Katsumata, and Katsufumi Sato.
\newblock Toward the quantification of a conceptual framework for movement
  ecology using circular statistical modeling.
\newblock \emph{{PLoS} {ONE}}, 7\penalty0 (11):\penalty0 e50309, nov 2012.
\newblock \doi{10.1371/journal.pone.0050309}.
\newblock URL \url{http://dx.doi.org/10.1371/journal.pone.0050309}.

\bibitem[Moreau et~al.(2012)Moreau, Fortin, Couturier, and
  Duchesne]{Moreau2012}
Guillaume Moreau, Daniel Fortin, Serge Couturier, and Thierry Duchesne.
\newblock Multi-level functional responses for wildlife conservation: the case
  of threatened caribou in managed boreal forests.
\newblock \emph{Journal of Applied Ecology}, 49\penalty0 (3):\penalty0
  611--620, may 2012.
\newblock \doi{10.1111/j.1365-2664.2012.02134.x}.
\newblock URL \url{http://dx.doi.org/10.1111/j.1365-2664.2012.02134.x}.

\bibitem[Rivest et~al.(2016)Rivest, Duchesne, Nicosia, and Fortin]{Rivest2015}
Louis-Paul Rivest, Thierry Duchesne, Aur{\'{e}}lien Nicosia, and Daniel Fortin.
\newblock A general angular regression model for the analysis of data on animal
  movement in ecology.
\newblock \emph{Journal of the Royal Statistical Society: Series C (Applied
  Statistics)}, pages n/a--n/a, oct 2016.
\newblock \doi{10.1111/rssc.12124}.
\newblock URL \url{http://dx.doi.org/10.1111/rssc.12124}.

\bibitem[Fryxell et~al.(2008)Fryxell, Hazell, Borger, Dalziel, Haydon, Morales,
  McIntosh, and Rosatte]{Fryxell2008}
J.~M. Fryxell, M.~Hazell, L.~Borger, B.~D. Dalziel, D.~T. Haydon, J.~M.
  Morales, T.~McIntosh, and R.~C. Rosatte.
\newblock Multiple movement modes by large herbivores at multiple
  spatiotemporal scales.
\newblock \emph{Proceedings of the National Academy of Sciences}, 105\penalty0
  (49):\penalty0 19114--19119, dec 2008.
\newblock \doi{10.1073/pnas.0801737105}.
\newblock URL \url{http://dx.doi.org/10.1073/pnas.0801737105}.

\bibitem[Langrock et~al.(2012)Langrock, King, Matthiopoulos, Thomas, Fortin,
  and Morales]{Langrock2012}
Roland Langrock, Ruth King, Jason Matthiopoulos, Len Thomas, Daniel Fortin, and
  Juan~M. Morales.
\newblock Flexible and practical modeling of animal telemetry data: hidden
  {M}arkov models and extensions.
\newblock \emph{Ecology}, 93\penalty0 (11):\penalty0 2336--2342, nov 2012.
\newblock \doi{10.1890/11-2241.1}.
\newblock URL \url{http://dx.doi.org/10.1890/11-2241.1}.

\bibitem[Leonard E.~Baum(1966)]{baum66}
Ted~Petrie Leonard E.~Baum.
\newblock Statistical inference for probabilistic functions of finite state
  {M}arkov chains.
\newblock \emph{The Annals of Mathematical Statistics}, 37\penalty0
  (6):\penalty0 1554--1563, 1966.
\newblock ISSN 00034851.
\newblock URL \url{http://www.jstor.org/stable/2238772}.

\bibitem[Morales et~al.(2004)Morales, Haydon, Frair, Holsinger, and
  Fryxell]{Morales2004}
Juan~Manuel Morales, Daniel~T. Haydon, Jacqui Frair, Kent~E. Holsinger, and
  John~M. Fryxell.
\newblock Extracting more out of relocation data: building movement models as
  mixtures of random walks.
\newblock \emph{Ecology}, 85\penalty0 (9):\penalty0 2436--2445, sep 2004.
\newblock \doi{10.1890/03-0269}.
\newblock URL \url{http://dx.doi.org/10.1890/03-0269}.

\bibitem[Holzmann et~al.(2006)Holzmann, Munk, Suster, and
  Zucchini]{Holzmann2006}
Hajo Holzmann, Axel Munk, Max Suster, and Walter Zucchini.
\newblock Hidden {M}arkov models for circular and linear-circular time series.
\newblock \emph{Environmental and Ecological Statistics}, 13\penalty0
  (3):\penalty0 325--347, sep 2006.
\newblock \doi{10.1007/s10651-006-0015-7}.
\newblock URL \url{http://dx.doi.org/10.1007/s10651-006-0015-7}.

\bibitem[Rydèn et~al.(2004)Rydèn, Moulines, and Douc]{Rydn2004}
Tobias Rydèn, Eric Moulines, and Randal Douc.
\newblock Asymptotic properties of the maximum likelihood estimator in
  autoregressive models with {M}arkov regime.
\newblock \emph{The Annals of Statistics}, 32\penalty0 (5):\penalty0
  2254--2304, oct 2004.
\newblock \doi{10.1214/009053604000000021}.
\newblock URL \url{http://dx.doi.org/10.1214/009053604000000021}.

\bibitem[Frühwirth-Schnatter(2013)]{Sylvia13}
Sylvia Frühwirth-Schnatter.
\newblock \emph{Finite Mixture and {M}arkov Switching Models (Springer Series
  in Statistics)}.
\newblock Springer, 2013.

\bibitem[Duchesne et~al.(2015)Duchesne, Fortin, and Rivest]{Duchesne2015}
Thierry Duchesne, Daniel Fortin, and Louis-Paul Rivest.
\newblock Equivalence between step selection functions and biased correlated
  random walks for statistical inference on animal movement.
\newblock \emph{{PLOS} {ONE}}, 10\penalty0 (4):\penalty0 e0122947, apr 2015.
\newblock \doi{10.1371/journal.pone.0122947}.
\newblock URL \url{http://dx.doi.org/10.1371/journal.pone.0122947}.

\bibitem[Mardia and Jupp(1999)]{Mardia00}
Kanti~V. Mardia and Peter~E. Jupp.
\newblock \emph{Directional Statistics}.
\newblock Wiley, 1999.
\newblock ISBN 0471953334.

\bibitem[Ailliot and Monbet(2012)]{ailliot2012}
Pierre Ailliot and Val{\'{e}}rie Monbet.
\newblock {M}arkov-switching autoregressive models for wind time series.
\newblock \emph{Environmental Modelling {\&} Software}, 30:\penalty0 92--101,
  apr 2012.
\newblock \doi{10.1016/j.envsoft.2011.10.011}.
\newblock URL \url{http://dx.doi.org/10.1016/j.envsoft.2011.10.011}.

\bibitem[Langrock and Zucchini(2011)]{Langrock2011}
R.~Langrock and W.~Zucchini.
\newblock Hidden {M}arkov models with arbitrary state dwell-time distributions.
\newblock \emph{Computational Statistics {\&} Data Analysis}, 55\penalty0
  (1):\penalty0 715--724, jan 2011.
\newblock \doi{10.1016/j.csda.2010.06.015}.
\newblock URL \url{http://dx.doi.org/10.1016/j.csda.2010.06.015}.

\bibitem[Hamilton(2008)]{hamilton2008regime}
James~D Hamilton.
\newblock Regime-switching models.
\newblock \emph{The new palgrave dictionary of economics}, 2, 2008.

\bibitem[Avriel(2003)]{avriel2003nonlinear}
Mordecai Avriel.
\newblock \emph{Nonlinear programming: analysis and methods}.
\newblock Courier Corporation, 2003.

\bibitem[Fortin et~al.(2013)Fortin, Buono, Fortin, Courbin, Gingras, Moorcroft,
  Courtois, and Dussault]{Fortin13}
Daniel Fortin, Pietro-Luciano Buono, André Fortin, Nicolas Courbin,
  Christian~Tye Gingras, Paul~R. Moorcroft, Réhaume Courtois, and Claude
  Dussault.
\newblock Movement responses of caribou to human-induced habitat edges lead to
  their aggregation near anthropogenic features.
\newblock \emph{The American Naturalist}, 181\penalty0 (6):\penalty0 827--836,
  2013.
\newblock ISSN 00030147, 15375323.
\newblock URL \url{http://www.jstor.org/stable/10.1086/670243}.

\bibitem[Bastille-Rousseau et~al.(2011)Bastille-Rousseau, Fortin, Dussault,
  Courtois, and Ouellet]{BastilleRousseau2011}
Guillaume Bastille-Rousseau, Daniel Fortin, Christian Dussault, R{\'{e}}haume
  Courtois, and Jean-Pierre Ouellet.
\newblock Foraging strategies by omnivores: are black bears actively searching
  for ungulate neonates or are they simply opportunistic predators?
\newblock \emph{Ecography}, 34\penalty0 (4):\penalty0 588--596, aug 2011.
\newblock \doi{10.1111/j.1600-0587.2010.06517.x}.
\newblock URL \url{http://dx.doi.org/10.1111/j.1600-0587.2010.06517.x}.

\bibitem[Dancose et~al.(2011)Dancose, Fortin, and Guo]{Dancose2011}
Karine Dancose, Daniel Fortin, and Xulin Guo.
\newblock Mechanisms of functional connectivity: the case of free-ranging bison
  in a forest landscape.
\newblock \emph{Ecological Applications}, 21\penalty0 (5):\penalty0 1871--1885,
  jul 2011.
\newblock \doi{10.1890/10-0779.1}.
\newblock URL \url{http://dx.doi.org/10.1890/10-0779.1}.

\bibitem[Courbin et~al.(2014)Courbin, Fortin, Dussault, and
  Courtois]{Courbin2014}
N.~Courbin, D.~Fortin, C.~Dussault, and R.~Courtois.
\newblock Logging-induced changes in habitat network connectivity shape
  behavioral interactions in the wolf{\textendash}caribou{\textendash}moose
  system.
\newblock \emph{Ecological Monographs}, 84\penalty0 (2):\penalty0 265--285, may
  2014.
\newblock \doi{10.1890/12-2118.1}.
\newblock URL \url{http://dx.doi.org/10.1890/12-2118.1}.

\bibitem[Smith(1974)]{smith}
James N.~M. Smith.
\newblock The food searching behaviour of two european thrushes i. description
  and analysis of search paths.
\newblock \emph{Behaviour}, 48\penalty0 (3/4):\penalty0 276--302, 1974.
\newblock ISSN 00057959.
\newblock URL \url{http://www.jstor.org/stable/4533575}.

\bibitem[Barraquand and Benhamou(2008)]{barraquand2008animal}
Fr{\'{e}}d{\'{e}}ric Barraquand and Simon Benhamou.
\newblock Animal movements in heterogeneous landscapes: identifying profitable
  places and homogeneous movement bouts.
\newblock \emph{Ecology}, 89\penalty0 (12):\penalty0 3336--3348, dec 2008.
\newblock \doi{10.1890/08-0162.1}.
\newblock URL \url{http://dx.doi.org/10.1890/08-0162.1}.

\bibitem[Patterson et~al.(2008)Patterson, Thomas, Wilcox, Ovaskainen, and
  Matthiopoulos]{PATTERSON2008}
T~Patterson, L~Thomas, C~Wilcox, O~Ovaskainen, and J~Matthiopoulos.
\newblock State{\textendash}space models of individual animal movement.
\newblock \emph{Trends in Ecology {\&} Evolution}, 23\penalty0 (2):\penalty0
  87--94, feb 2008.
\newblock \doi{10.1016/j.tree.2007.10.009}.
\newblock URL \url{http://dx.doi.org/10.1016/j.tree.2007.10.009}.

\bibitem[Albertsen et~al.(2015)Albertsen, Whoriskey, Yurkowski, Nielsen, and
  Flemming]{Albertsen2015}
Christoffer~Moesgaard Albertsen, Kim Whoriskey, David Yurkowski, Anders
  Nielsen, and Joanna~Mills Flemming.
\newblock Fast fitting of non-{G}aussian state-space models to animal movement
  data via template model builder.
\newblock \emph{Ecology}, 96\penalty0 (10):\penalty0 2598--2604, oct 2015.
\newblock \doi{10.1890/14-2101.1}.
\newblock URL \url{http://dx.doi.org/10.1890/14-2101.1}.

\bibitem[Basille et~al.(2012)Basille, Fortin, Dussault, Ouellet, and
  Courtois]{Basille2012}
Mathieu Basille, Daniel Fortin, Christian Dussault, Jean-Pierre Ouellet, and
  R{\'{e}}haume Courtois.
\newblock Ecologically based definition of seasons clarifies predator-prey
  interactions.
\newblock \emph{Ecography}, 36\penalty0 (2):\penalty0 220--229, apr 2012.
\newblock \doi{10.1111/j.1600-0587.2011.07367.x}.
\newblock URL \url{http://dx.doi.org/10.1111/j.1600-0587.2011.07367.x}.

\bibitem[Ingrassia and Rocci(2011)]{Ingrassia2011}
Salvatore Ingrassia and Roberto Rocci.
\newblock Degeneracy of the {EM} algorithm for the {MLE} of multivariate
  {G}aussian mixtures and dynamic constraints.
\newblock \emph{Computational Statistics {\&} Data Analysis}, 55\penalty0
  (4):\penalty0 1715--1725, apr 2011.
\newblock \doi{10.1016/j.csda.2010.10.026}.
\newblock URL \url{http://dx.doi.org/10.1016/j.csda.2010.10.026}.

\bibitem[Biernacki et~al.(2003)Biernacki, Celeux, and Govaert]{Biernacki2003}
Christophe Biernacki, Gilles Celeux, and G{\'{e}}rard Govaert.
\newblock Choosing starting values for the {EM} algorithm for getting the
  highest likelihood in multivariate {G}aussian mixture models.
\newblock \emph{Computational Statistics {\&} Data Analysis}, 41\penalty0
  (3-4):\penalty0 561--575, jan 2003.
\newblock \doi{10.1016/s0167-9473(02)00163-9}.
\newblock URL \url{http://dx.doi.org/10.1016/S0167-9473(02)00163-9}.

\end{thebibliography}
%

%

%
%
%

\appendix

\section{Filtering-Smoothing Algorithm for the Markov Specification}
\label{a:A}
In the E-step of the ($s$+1)-th iteration of the EM algorithm we have to compute two posterior expectations involving the hidden $S_{kt}$, $k=1,\ldots,K$, $t=0,\ldots,T$, conditionally on the observed data  $\mathcal{F}_{T}^o$:
\begin{eqnarray}
\label{E1}
\mathbb{E}( S_{kt}|\mathcal{F}_{T}^o,\hat{\mathbf{\theta}}_s)&=&\mathbb{P}( S_{kt}=1|\mathcal{F}_{T}^o,\hat{\mathbf{\theta}}_s)\\
\mathbb{E} (S_{h,t-1} S_{k,t}|\mathcal{F}_{T}^o,\hat{\mathbf{\theta}}_s) 
&=&\mathbb{P}(S_{h,t-1}=1|S_{kt}=1,\mathcal{F}_{T}^o,\hat{\mathbf{\theta}}_s)\mathbb{P}(S_{kt}=1 |\mathcal{F}_{T}^o,\hat{\mathbf{\theta}}_s), \label{E3}
\end{eqnarray}
where $\hat{\mathbf{\theta}}_s$ is the maximized vector of parameters after the $s$-th step of the EM algorithm. The first probability on the RHS of (\ref{E3}) can be computed with Bayes' theorem because, as we can see from Figure \ref{f:fig1}
, $S_{t-1}$ is independent of the observed data from time $t$ to $T$ (i.e. $\lbrace \mathcal{F}_{t+s}^o \rbrace_{s \geq 0} \setminus \mathcal{F}_{t-1}^o$)  given $S_t$ and $\mathcal{F}_{t-1}^o $ :
\begin{eqnarray*}
\mathbb{P}(S_{h,t-1}=1|S_{kt}=1,\mathcal{F}_{T}^o,\hat{\mathbf{\theta}}_s)=\frac{\hat{\pi}_{hk}^{(s)} \mathbb{P}(S_{h,t-1}=1 |\mathcal{F}_{t-1}^o,\hat{\mathbf{\theta}}_s)}{\sum_{j=1}^K \hat{\pi}_{jk}^{(s)} \mathbb{P}(S_{j,t-1}=1 |\mathcal{F}_{t-1}^o,\hat{\mathbf{\theta}}_s) } ,\ k=1,\ldots,K,\ t=0,\ldots,T.
\end{eqnarray*}
Finally, to compute the remaining conditional probabilities in the posterior expectations \eqref{E1} and \eqref{E3}, we adapt the classical filtering-smoothing algorithm of \cite{Sylvia13}.
\\
\rule{\linewidth}{.5pt}
\\
\textsc{Filtering-smoothing algorithm to implement the E-step \\ of the  $(s+1)$-th iteration of the EM algorithm.}
\\
\rule{\linewidth}{.5pt}
\\
\begin{itemize}
\item[\textbf{Filter}] Compute $\mathbb{P}(S_{tl}=1|\mathcal{F}_{t}^o,\hat{\mathbf{\theta}}_s)$, for every $l=1,\dots,K$ :
$$
\mathbb{P}(S_{lt}=1|\mathcal{F}_{t}^o,\hat{\mathbf{\theta}}_s)=\frac{f_l(y_t|\mathcal{F}_{t-1}^o,\hat{\mathbf{\theta}}_s) g_l(d_t,\hat{\mathbf{\theta}}_s)\mathbb{P}(S_{tl}=1|\mathcal{F}_{t-1}^o,\hat{\mathbf{\theta}}_s)}{\sum_{k=1}^K f_k(y_t|\mathcal{F}_{t-1}^o,\hat{\mathbf{\theta}}_s) g_k(d_t,\hat{\mathbf{\theta}}_s) \mathbb{P}(S_{kt}=1|\mathcal{F}_{t-1}^o,\hat{\mathbf{\theta}}_s)},
$$
where $\mathbb{P}(S_{l,1}=1|\mathcal{F}_{0}^o,\hat{\mathbf{\theta}}_s)=\sum_{k=1}^K \hat{\pi}_{kl}^{(s)}({\pi}_0)_k$ \\ and  
\begin{equation}
\label{a:Predictive}  
\mathbb{P}(S_{lt}=1|\mathcal{F}_{t-1}^o,\hat{\mathbf{\theta}}_s)=\sum_{k=1}^K \hat{\pi}_{kl}^{(s)}\mathbb{P}(S_{k,t-1}=1|\mathcal{F}_{t-1}^o,\hat{\mathbf{\theta}}_s) ,
\end{equation}
for $t=2,\dots,T.$
\item[\textbf{Smooth}] Compute $\mathbb{P}(S_{lt}=1|\mathcal{F}_{T}^o,\hat{\mathbf{\theta}}_s)$, for every $l=1,\dots,K$:
\begin{itemize}
\item[\textsf{S-step 1}] For $t=T$, set $\mathbb{P}(S_{lT}=1|\mathcal{F}_{T}^o,\hat{\mathbf{\theta}}_s)$, the conditional probability computed
at the last filtering step.

\item[\textsf{S-step 2}] Recursion: For $t=T-1,\ldots,0$, compute:
\begin{equation}
\label{a:smooth}
\mathbb{P}(S_{lt}=1|\mathcal{F}_{T}^o,\hat{\mathbf{\theta}}_s)=\sum_{k=1}^K \frac{\hat{\pi}_{lk}^{(s)}\mathbb{P}(S_{lt}=1|\mathcal{F}_{t}^o,\hat{\mathbf{\theta}}_s) \mathbb{P}(S_{k,t+1}=1|\mathcal{F}_{T}^o,\hat{\mathbf{\theta}}_s)}{\sum_{j=1}^K \hat{\pi}_{jk}^{(s)} \mathbb{P}(S_{jt}=1|\mathcal{F}_{t}^o,\hat{\mathbf{\theta}}_s)}.
\end{equation}

\end{itemize}

\end{itemize}

\rule{\linewidth}{.5pt}

\section{Markov Approximation of a Semi-Markov Process}
\label{a:B}
We follow the idea of approximating a semi-Markov process by an extended Markov chain as introduced by \cite{Langrock2011}. We present here the details for a two state process.

Let us suppose that $S_{0:T}$ is a two state semi-Markov process with $Q_1$ and $Q_2$ the probability mass functions of the dwell times in both states. Then using notation of Section~\ref{ss:Markov}, $n_h,q_{h}$ denote the parameters (size and probability) of the dwell time distribution in state $h$, $Q_h$.  The stochastic
 behavior of this process is approximated by a Markov chain $\tilde{S}_{0:T}$  with state space  $\lbrace (i,k),i=1,2,k=1,\ldots, m_i\rbrace$, where $i$ still denotes the state of the animal and $k$ is the dwell time in this state. Given that $\tilde{S}_t=(i, k)$, two moves are possible: one is to go to state $(3-i,1)$ with probability 
 $$
 \pi_{(i, k)(3-i,1)}(n_i,q_{i})=Q_i(k)/\{Q_i(k)+Q_i(k+1)+\dots\},
 $$
  or to state $(i, \min(m_i,k+1))$ with probability $1- \pi_{(i, k)(3-i,1)}(n_i,q_{i})$. In applications the $Q_i$ are often chosen to be the probability mass functions of negative binomial distributions and in this case, typical values of the $m_i$'s are around 30.

Figure B.4 is an example of the transition matrix contructed as exposed with two states and $m_1=4$, $m_2=4$, it means that the enlarged Markov chain $\tilde{S}_{0,T}$ is an 7 state Markov chain. We also give the basic example of $m_1=m_2=1$, which reduces to the classical Markov chain model:
$$
\bordermatrix{
    &(1,1) & (2,1) \cr
(1,1) &  1-\pi_{(1, 1)(2,1)}(1,q_{1})  &  \pi_{(1, 1)(2,1)}(1,q_{1})  \cr
(2,1) &  \pi_{(2, 1)(1,1)}(1,q_{2}) & 1-\pi_{(2, 1)(2,2)}(1,q_{2})\cr
} = \bordermatrix{
    &(1,1) & (2,1) \cr
(1,1) &  1-q_{1} &  q_{1}  \cr
(2,1) &  q_{2}& 1-q_{2}\cr
} .
$$
%
\begin{landscape}
\begin{small}
\begin{figure}[h] 
\centering
\bordermatrix{
    &(1,1) & (1,2) &(1,3) & (1,4) & (2,1) & (2,2) &(2,3) \cr
(1,1) &  0 & 1-\pi_{(1, 1)(2,1)}(n_1,q_{1})  & 0 & 0 &  \pi_{(1, 1)(2,1)}(n_1,q_{1}) & 0 & 0  \cr
(1,2)&  0 & 0 & 1-\pi_{(1, 2)(2,1)}(n_1,q_{1})  & 0 & \pi_{(1, 2)(2,1)}(n_1,q_{1})  & 0 & 0  \cr
(1,3)&  0 & 0 & 0 & 1-\pi_{(1, 3)(2,1)}(n_1,q_{1})  &  \pi_{(1, 3)(2,1)}(n_1,q_{1})  & 0 & 0 \cr
(1,4) &  0 & 0 & 0 & 1-\pi_{(1, 4)(2,1)}(n_1,q_{1}) &  \pi_{(1, 4)(2,1)}(n_1,q_{1}) & 0 & 0  \cr
(2,1) &  \pi_{(2, 1)(1,1)}(n_2,q_{2}) & 0 & 0 & 0 & 0 & 1-\pi_{(2, 1)(2,2)}(n_2,q_{2})& 0 \cr
(2,2)&   \pi_{(2, 2)(1,1)}(n_2,q_{2}) & 0 & 0 & 0 & 0& 0  &  1- \pi_{(2, 2)(1,1)}(n_2,q_{2})\cr
(2,3)&  \pi_{(2, 3)(1,1)}(n_2,q_{2}) & 0 & 0 & 0& 0  & 0 &1- \pi_{(2, 3)(1,1)}(n_2,q_{2}) \cr
}
\caption{Example of transition matrix with $K=2$ states, $m_1=4$ and $m_2=3$.}
\end{figure}
\end{small}
\label{f:semi}
\end{landscape}

\section{Numerical Details}
%

\subsection{Finding the Global Maximum of the Likelihood Function}
\label{a:C1}
Due to the complexity of the model, the EM algorithm may converge to local or spurious maxima of the likelihood function. We have observed that the EM algorithm can quickly converge to spurious maxima (in less than 10 iterations). This is due to the fact that for some parameters in $\mathbf{\kappa}$, $\mathbf{\lambda}$ and $\mathbf{\pi}$, the likelihood is unbounded, a common phenomenon in the case of mixture models \cite{Ingrassia2011}.

 To deal with this, we run the EM algorithm with many random starting values for a few iterations and check for spurious and local solutions. We then choose the parameter values that yield the highest likelihood as the starting point of a new EM algorithm that we run until convergence. This strategy of combining short- and long-run EM algorithms to avoid possible local and spurious maxima is known as the 1em-EM algorithm (\cite{Biernacki2003}). To obtain an estimate of the variance matrix of the maximum likelihood estimator we run one iteration of the quasi-Newton algorithm to obtain the value of the inverse of the Hessian matrix of the observed log-likelihood function evaluated at the maximum likelihood estimates. Here is an algorithmic description of this procedure.
\\
\rule{\linewidth}{.5pt}
\\
\textsc{Finding the global maximum of the observed log-likelihood function}
\\
\rule{\linewidth}{.5pt}
\\
\textbf{Preliminary step} :
\begin{itemize}

\item Let $\mathbf{\theta}_1,\dots,\mathbf{\theta}_N$ be $N$ random initial starting values. (In our application of this method, we chose $N=50$.)

\item For $i=1,\dots,N$, run the EM algorithm until the first of (i) 50 iterations or (ii) the greatest relative difference in parameter value between successive iterations is less than 1\%. Denote the
estimators obtained at the end of this step $\hat{\mathbf{\theta}}_i$, $i=1,\ldots,N$.
\end{itemize}

\textbf{Avoid spurious maxima : }

\begin{itemize}

\item For each $\hat{\mathbf{\theta}}_i$, $i=1,\ldots,N$, compute the stationary distribution of the Markov chain, $\hat{\nu}^{(i)}_k$, $k=1,\dots,K$.

\item Only keep the $\{\hat{\mathbf{\theta}}_i,i\in I\}$ such that
$$\min_{k=1,\dots,K} \hat{\nu}^{(i)}_k>\epsilon\ \ \ \ \mbox{and}\ \ \ \
\max_{j=1,\ldots,p;k=1,\ldots,K} |{\kappa}^{(k)(i)}_j|<M.$$
(In our application of this method, we chose $\epsilon=0.001$ and $M=100$.)
\end{itemize}

\textbf{Avoid local maxima : }

\begin{itemize}

\item Put $\mathbf{\theta}_0= \text{arg}\max_{i \in I} \text{L}(\hat{\mathbf{\theta}}_i)$.

\end{itemize}

\textbf{Long-run EM algorithm} :

\begin{itemize}

 \item Start the EM algorithm at $\mathbf{\theta}_{0}$ and run it until the first of (i) 10~000 iterations or (ii) the greatest relative difference in parameter value between successive iterations is less than $10^{-8}$.
 \end{itemize}

\textbf{Quasi-Newton iteration} :

\begin{itemize}

\item Run one iteration of the quasi-Newton algorithm with the output of the long-run EM algorithm as initial value to get the final global maximum likelihood estimators of the model parameters and an estimation of their variance matrix.

\end{itemize}
\rule{\linewidth}{.5pt}

\subsection{A Note on the Initial Distribution $(\pi_0)_k,k=1,\ldots,K$}
\label{a:C2}
Calculation of either the observed or complete data likelihood involves the initial distribution of the Markov chain, $(\pi_0)_k$, $k=1,\ldots,K$.
We have decided to fit the model twice. For the first fit we use $(\pi_0)_k=1/K$, $k=1,\ldots,K$, the uniform distribution over the $K$ states. Then we fit the model again, but this time with $(\pi_0)_k=\hat{\nu}_k$, the stationary distribution of the chain computed from the first model fit.

\subsection{A Note on the Identifiability of the Model up to State Label Switching}
\label{a:C3}
We can easily see that the value of the likelihood function remains the same if we relabel the states. We therefore define the states at the end of each M-step as follows: we give label $i$ to the state with the $i$-th smallest $\hat{\kappa}_0^{(k)}$.
\subsection{Multi modality of the log-likelihood function with homogeneous errors}
\label{a:C4}
In this section we want to illustrate the multi modality of the log-likelihood of the biased correlated random walk model with homogeneous errors in a very simple example. Let us suppose that the Markov chain has $K=2$ states and define the vector
\begin{equation}\label{V.ex}
\mathbf{V}_t^{(k)}=1 \times \left(\begin{array}{c}
\cos(y_{t-1})\\
\sin(y_{t-1})
\end{array}  \right)+\beta^{(k)}  \left(\begin{array}{c}
\cos(x_{t})\\
\sin(x_{t})
\end{array} \ \right) , t=1,\ldots,T,
\end{equation}
where $x_t$ is uniformly distributed on the unit circle, $\beta^{(1)}=-0.2$, $\beta^{(2)}=0.7$ and let us denote $\mu_t^{(k)}$ and $\ell_t^{(k)}$ the direction and the length of the vector (\ref{V.ex}). The direction $y_t$ is simulated as a von Mises distribution with mean direction $\mu_t^{(k)}$ and concentration $\kappa^{(k)}$, with $\kappa^{(1)}=0.1$, $\kappa^{(2)}=0.3$ .

\begin{figure}[H]
\centering
      \includegraphics[scale=0.7]{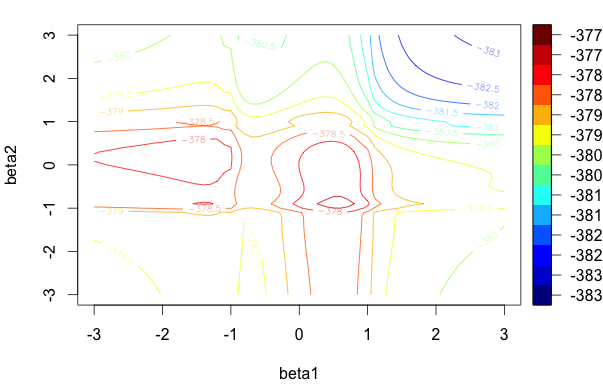}
         \caption{The log-likelihood function of the parameters $(\beta^{(1)} ,\beta^{(2)} )$ with sample size $n=50$.}
         \label{f:deuxLikelihood}
\end{figure}

The distances are simulated using different gamma distributions in each state ($\lambda_1^{(1)}=0.5$, $\lambda_2^{(1)}=2$ and $\lambda_1^{(2)}=1$, $\lambda_2^{(2)}=0.2$) and the Markov chain’s parameters are $q_1=0.1$ and $q_2=0.2$.  All the parameters are assumed to be known except for $\beta_1$ and $\beta_2$.  Figure \ref{f:deuxLikelihood} gives contour plots of the log-likelihood for $ (\beta_1,\beta_2)$ obtained with a simulated sample of size $n=50$.  It shows two local maxima, around $(\beta_1,\beta_2)$ equal to $(-1.5,0)$ and $(-1,.5)$. Such a multimodal log-likelihood was obtained 4 times out of 20 simulations.


For standard angular regression problems, the concensus model has a concave log-likelihood, see Section \ref{sec:Obsprocess}, that is easily maximized. We expect that these good numerical properties will also apply to the general random walk directional model introduced in Section \ref{sec:Obsprocess}. 
 Actually, such a unimodal log-likelihood was obtained 20 times out of 20 simulations.

%

\subsection{Characteristic Trajectory Under First Scenario of  Simulation}
\label{a:C5}
\begin{figure}[H]
\centering
      \includegraphics[scale=0.7]{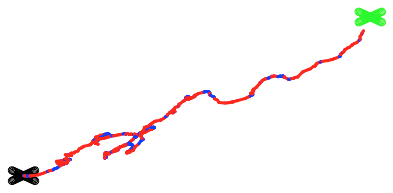}
         \caption{Characteristic trajectory of one simulation under scenario 1 in Table \ref{t:scenario}. The first state ($k=1$) is coloured in red, while the second state ($k=2$) is blue. The black cross is the starting point and the green cross is the target.}
\end{figure}

\subsection{Validation}
\label{a:valid}
In order to compare the observed trajectory to trajectories simulated from the fitted model, we first notice that the animal has two symetrical sub-trajectories. In the first one, the animal is going from the departure (red circle in Figure~\ref{f:trajobs}) to the bottom left corner of Figure~\ref{f:trajobs}. In the second one, the animal returns to a final point (blue circle in Figure~\ref{f:trajobs})) that is close to the starting point of the first sub-trajectory.
This symetry means that we can save computing time by only comparing trajectories simulated by the model to one of the two observed sub-trajectories; we chose to compare them to the first observed sub-trajectory.

To make the computational load manageable, we simulate trajectories using a slightly modified model:
\begin{itemize}

\item we drop the variable $x_{\text{cut}}$ since it would take too much time to recompute at each simulated time step and its effect in the model is weak;

\item we simplify the variable $x_{\text{center}}$, which would also take time to recompute, and redefine it as the direction to the centroid of the locations in the bottom left corner of Figure~\ref{f:trajobs}.

\end{itemize}

Except for these simplifications the trajectories are simulated using the estimated parameters in Table~\ref{t:resultCar}. We simulate $N=500$ trajectories to produce the following figures and statistics. Figure~\ref{fig:cum} and Figure~\ref{fig:hist} depict, respectively, the empirical cumulative distribution function and histogram of the number of time steps required to return to circle of diameter 2 km around of the centroid of the locations in the bottom left corner. We stop the trajectory after 10,000 steps if it does not reach the neighborhood (this happened only in 9 of the 500 simulations, so less than 2\% of the times.

\begin{figure}[h!]
\begin{minipage}[c]{.45\linewidth}
\begin{center}
\includegraphics[scale=0.45]{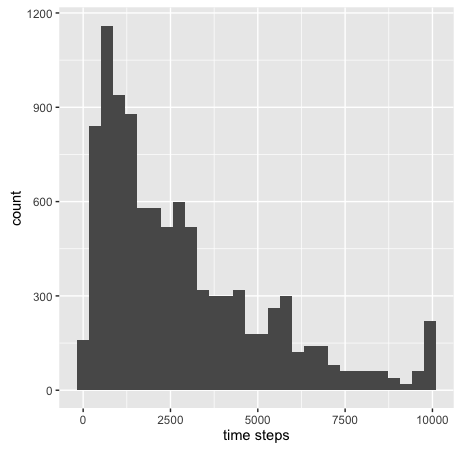}
\caption{Histogram of the number of time steps needed for the simulated trajectories to reach a neighborhood of the centroid of the locations in the bottom left corner of the map. }
\label{fig:hist}
\end{center}
\end{minipage}
\hfill
\begin{minipage}[c]{.45\linewidth}
\begin{center}
\includegraphics[scale=0.45]{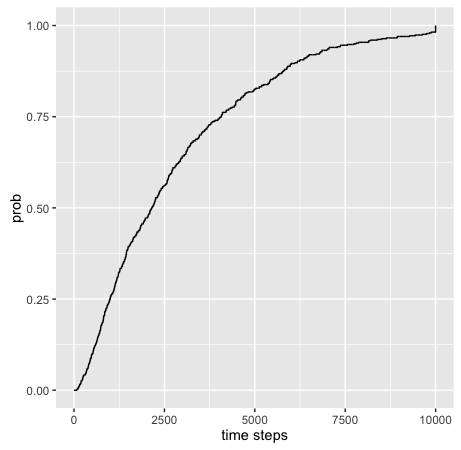}
\caption{Cumulative of the number of time steps needed for the simulated trajectories to reach a neighborhood of the centroid of the locations in the bottom left corner of the map.}
\label{fig:cum}
\end{center}
\end{minipage}
\end{figure}
Table \ref{tab:estimprob} gives estimates of the probabilities that the animal arrives in a neighborhood of the centroid of the locations in the bottom left corner of the map based on the 500 simulations.
\begin{table}[H]
\centering
\caption{Quantiles of the return's time steps distribution.}\label{tab:estimprob}
\begin{tabular}{ccccccc}
\hline
time steps  & 500 & 1000 & 2000 & 5000 &  10000\\
\hline
\hline
probability & 0.098 &  0.25& 0.472& 0.826 & 0.982 \\
\hline
\end{tabular}
\label{summarydistance1}
\end{table}

Finally, we show some randomly chosen trajectories from the simulation model. The color corresponds to the same state presented in the analysis of the caribou data in Section~\ref{s:caribou} i.e., red for the exploratory state and blue for the encamped one.

\begin{figure}[h!]
\begin{minipage}[c]{.45\linewidth}
\begin{center}
\includegraphics[scale=0.3]{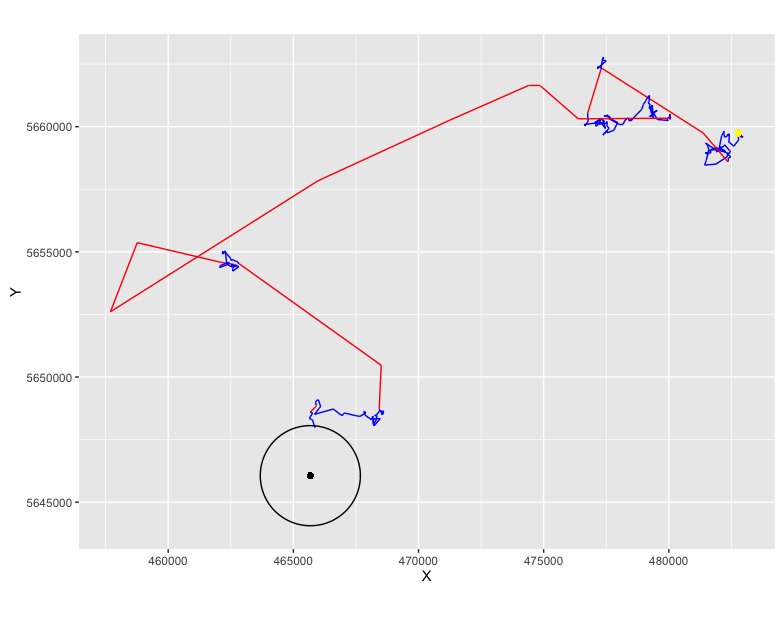}
\end{center}
\end{minipage}
\hfill
\begin{minipage}[c]{.45\linewidth}
\begin{center}
\includegraphics[scale=0.3]{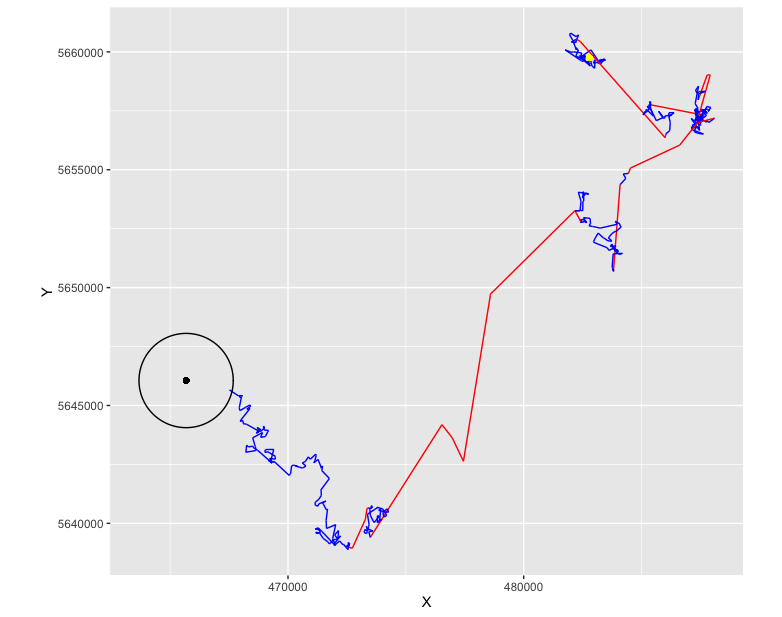}
\end{center}
\end{minipage}
\end{figure}

\begin{figure}[h!]
\begin{minipage}[c]{.45\linewidth}
\begin{center}
\includegraphics[scale=0.3]{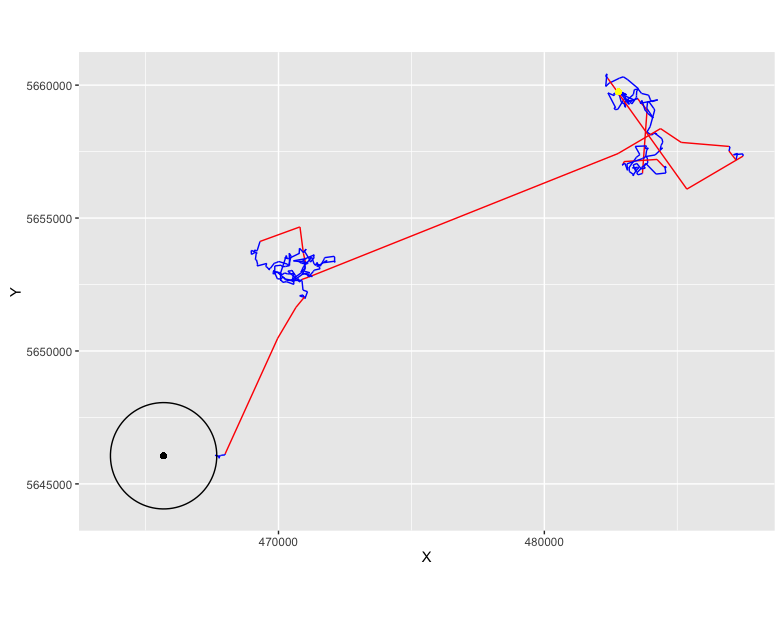}
\end{center}
\end{minipage}
\hfill
\begin{minipage}[c]{.45\linewidth}
\begin{center}
\includegraphics[scale=0.3]{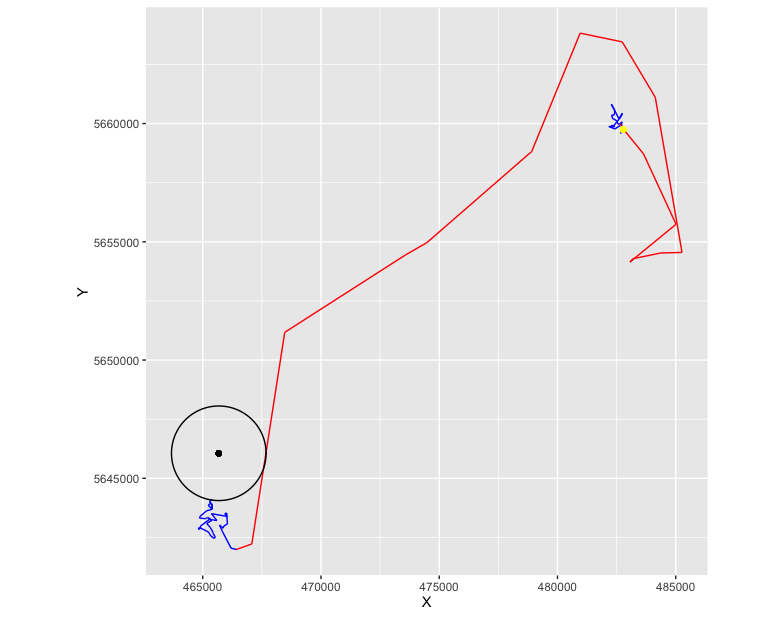}
\end{center}
\end{minipage}
\end{figure}

\begin{figure}[h!]
\begin{minipage}[c]{.45\linewidth}
\begin{center}
\includegraphics[scale=0.25]{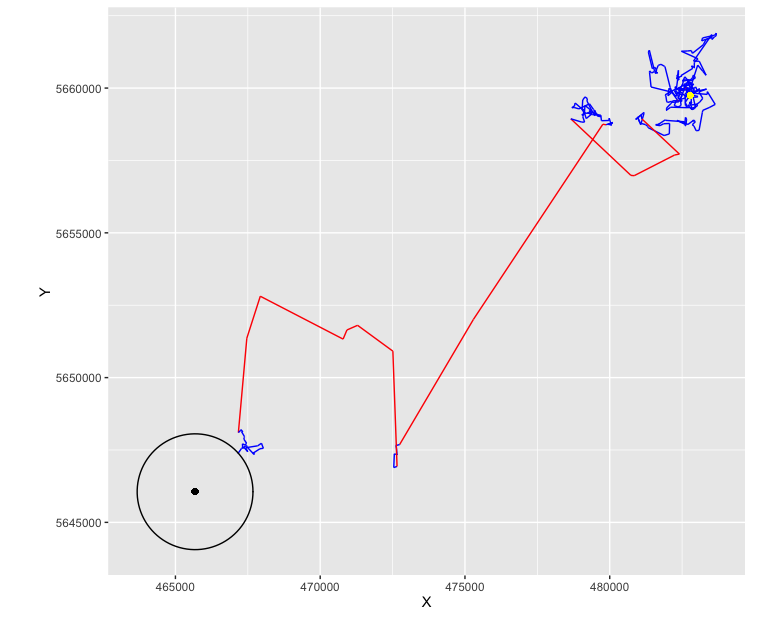}
\end{center}
\end{minipage}
\hfill
\begin{minipage}[c]{.45\linewidth}
\begin{center}
\includegraphics[scale=0.25]{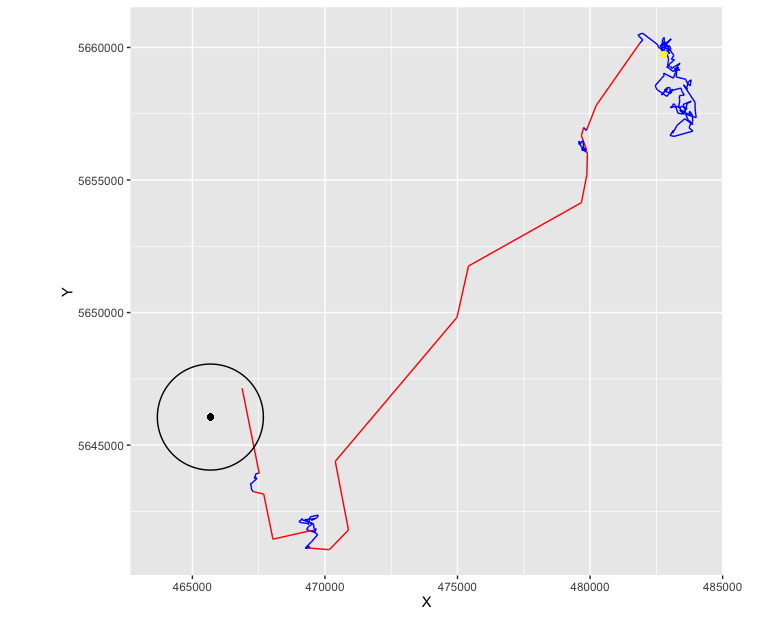}
\end{center}
\end{minipage}
\end{figure}

\newpage
\newpage

\section{Descriptive Statistics  on the Trajectory of the Caribou}
\label{a:D}
We present an exploratory analysis of the trajectory data of the caribou. These data consist of 617 observations of the form $(y_t,d_t,\mathbf{x}_t)$, where  $y_t$ and $d_t$ respectively represent the direction (bearing) and the distance between the caribou location at time step $t$ and time step $t+1$ and $\mathbf{x}_t$ are the values of five exploratory angle variables. Besides directional persistence, $y_{t-1}$, several explanatory angles were considered in the analysis.  The ones kept for the final model are $x_{\text{cut}}$, the direction to the closest regenerating wood cut (that is an area where wood had been cut between 5 and 20 years ago), and the direction to the centroid of the closest patch $x_{\text{centre} }$. A patch is an aggregation of locations recently visited by the animal;  at a given step, the center of gravity of the patch's locations is recalculated and $x_{\text{centre} }$ is the angle of the segment joining the animal's position and the updated center of gravity;  see the Appendix of \cite{Latombe2014} for more details. In ecology, such a variable is called spatial memory effect, because animals tend to come
back to familiar areas for forage and safety.

\subsection{Observed Trajectory}
\begin{figure}[H]
\centering
\includegraphics[scale=0.5]{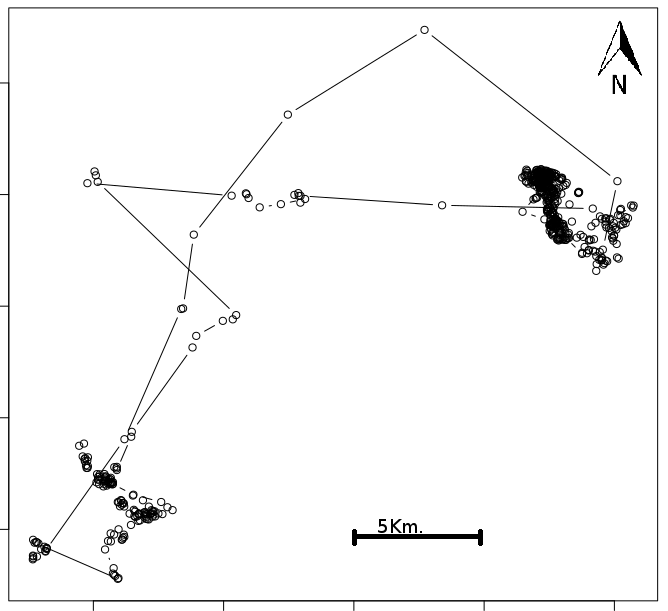}
\caption{Caribou trajectory from December 28 2005 to April 15 2006 .}
\label{f:trajectoire1}
\end{figure}
Figure~\ref{f:trajectoire1} shows the trajectory of the caribou. It strongly suggests two different movement behaviours: an ``encamped" state characterized by short traveled distances between two steps and an ``exploratory" state with longer traveled distances.

\subsection{Partitioning the Data According to the Distance Traveled}
To assess whether the directionality of the steps is different between the ``encamped'' and the ``exploratory'' states, we want to partition the dataset into two subsets: one with `` short" traveled distances and one with ``long" traveled distances. The cut point between ``short" and ``long" distances has to be precisely defined.
\begin{figure}[H]
\centering
\includegraphics[scale=0.3]{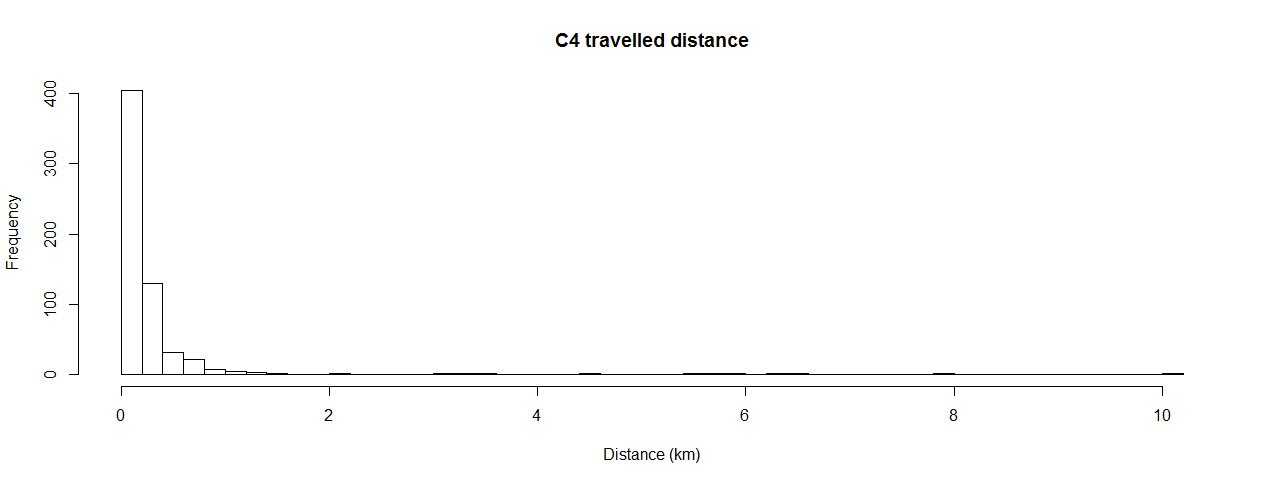}
\caption{ Histogram of travelled distances of the caribou from December 28 2005 to April 15 2006. }
\label{f:distance1}
\end{figure}
\begin{table}[H]
\centering
\caption{Summary statistics of the travelled distance (km).}
\begin{tabular}{ccccccc}
\hline
$n$ & Min & $Q_1$ & Median & Mean & $Q_3$ & Max  \\
\hline
\hline
617 & 0.0015 & 0.0642 & 0.1259 & 0.312 & 0.249 & 10.04 \\
\hline
\end{tabular}
\label{summarydistance1}
\end{table}
The histogram shown in Figure \ref{f:distance1} and the summary statistics from Table \ref{summarydistance1} indicate that most of the traveled distances are ``short", but with a non-negligible proportion of much longer distances.
We therefore fit a mixture of exponential distributions to the sample of traveled distances and obtain the results given in Table \ref{tabA2}.
\begin{table}[H]
\centering
\caption{Fitted mixture of exponential distribution on the distance. For $i=1,2$, $\pi_i$ and $\lambda_i$ respectively represent the weight and the mean of the exponential distribution $i$. }\label{tabA2}
\begin{tabular}{cccc}
\hline
$\pi_1$ & $\lambda_1$ & $\pi_2$ & $\lambda_2$  \\
\hline
\hline
 0.09 & 0.4602&0.91&7.3119\\
\hline
\end{tabular}

\end{table}
Finally, we can define a critical distance (the cutoff point between the ``encamped'' and the ``exploratory'' states), $d_{\text{Critical}}$, as the weighted mean of the two exponential distributions in the mixture model:
$$
d_{\text{critical}}=\frac{{\pi_1}}{{\lambda_1}}+\frac{{\pi_2}}{{\lambda_2}} \approx 0.3120 \text{			km.}
$$
This partitions the dataset into two subsets: one subset with 500 observations with distances traveled less than $d_{\text{Critical}}$ and one subset with 117 observations with traveled distances greater than $d_{\text{Critical}}$.

\subsection{Directionality of Movement Conditional on Distance Traveled}
We can now analyze the distribution of the turning angles ($y_t-y_{t-1},t=1,\ldots,T$) in the two subsets of the dataset defined by $d_{\text{Critical}}$.
\renewcommand{\arraystretch}{1}
\begin{table}[H]
\centering
\caption{Exploratory analysis of the distribution of the turning angles in each subset of the data defined by $d_{\text{Critical}}$.}
\begin{tabular}{ccc}
\hline
  & Distance traveled $< d_{\text{critical}}$ & Distance traveled $> d_{\text{critical}}$  \\
\hline
\hline
 Circular histogram &
 \includegraphics[scale=0.2]{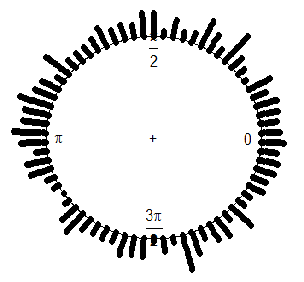}  &\includegraphics[scale=0.2]{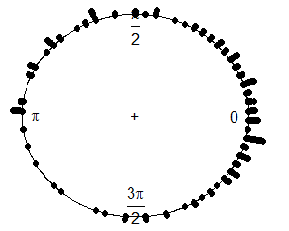}
 \\
\hline
Subset size & 500 & 117 \\
\hline
Mean direction (Rad) & 1.933 & 0.40 \\
\hline
Resultant length & 0.0253 & 0.3199 \\
\hline
$p_{\text{value}}$, Kuiper's test of uniformity  & $>0.15$ & $<0.01$ \\
\hline
\end{tabular}
\label{TAcaribou1}
\end{table}
Table \ref{TAcaribou1} shows that when the distance traveled exceeds $d_{\text{Critical}}$,
the animal path tends to be a concatenation of straight lines, as the turning angles are not uniformly distributed over the circle, but rather concentrated around direction 0. But when the distance traveled is less
than $d_{\text{Critical}}$, uniformity of the distribution of the turning angles is not ruled out, implying that the animal tends to turn around and does not have a preferred direction.

The final step of our exploratory analysis is to see if the environmental targets have the same influence on the animal's movement in the ``encamped'' and ``exploratory'' states.
To do so, we have fitted to the entire dataset our proposed model with $K=1$ state and the vector in the directional model (see (7) in the main paper) as
\begin{eqnarray}
\nonumber\mathbf{V}_t &=& \kappa_0 \left(\begin{array}{c}
\cos(y_{t-1})\\
\sin(y_{t-1})
\end{array}  \right)+ \kappa_1 \left(\begin{array}{c}
\cos(x_{\text{cut}})\\
\sin(x_{\text{cut}})
\end{array} \ \right) +\kappa_2 \left(\begin{array}{c}
\cos(x_{\text{center}})\\
\sin(x_{\text{center}})
\end{array} \ \right)  \\
&+& \nonumber \kappa_3 \times z_t \left(\begin{array}{c}
\cos(y_{t-1})\\
\sin(y_{t-1})
\end{array}  \right)+ \kappa_4 \times z_t \left(\begin{array}{c}
\cos(x_{\text{cut}})\\
\sin(x_{\text{cut}})
\end{array} \ \right) +\kappa_5 \times z_t \left(\begin{array}{c}
\cos(x_{\text{center}})\\
\sin(x_{\text{center}})
\end{array} \ \right)  ,\\\label{modelVt}
\end{eqnarray}
with $z_t$ take on value 1 if ${d_t>d_{\text{critical}}}$ and value 0 otherwise, for $t=0,\ldots,T$. After a backward selection of the variables based on Wald tests with significance levels of 0.05, we have obtained the
model summarized in Table \ref{onestate}. These results agree with the observations from Table \ref{TAcaribou1} and suggest that the directionality of the movement of the animal is different in the
two subsets of the data defined by $d_{\text{Critical}}$.
\begin{table}[H]
\centering
\caption{Parameter estimates of the final model for the directional model with directional mean vector given by (\ref{modelVt}) fitted to the entire dataset.}\label{onestate}
\begin{tabular}{ccccc}
\hline
\hline
 &$\kappa_0 $& $\kappa_1 $ & $\kappa_2 $ &$\kappa_3 $\\
\hline
\hline
Estimate & 0.0300 & 0.1457 & 0.2687 & 0.4176 \\
\hline
\hline
S.e. & 0.0635 &0.0642 & 0.0635 & 0.1515 \\
\hline
\end{tabular}
\end{table}

\label{lastpage}
%
%

%
%
%
%
%
%
%

\label{lastpage}
\end{document}